\documentclass[times,twocolumn,final]{elsarticle}

\usepackage{medima}
\usepackage{framed,multirow}
\usepackage{soul}
\usepackage{amssymb}
\usepackage{latexsym}
\usepackage{stackengine}

\usepackage{url}
\usepackage{xcolor}
\usepackage{todonotes}
\usepackage{hyperref}
\usepackage{dblfloatfix}
\usepackage{multirow}
\usepackage{amsmath}
\usepackage{textcomp}
\usepackage{siunitx}
\DeclareMathOperator{\E}{\mathbb{E}}

\definecolor{newcolor}{rgb}{.8,.349,.1}

\tikzset{/tikz/notestyleraw/.append style={text=black!70!white}}
\journal{Arxiv}

\begin{document}

\verso{Tian Xia \textit{et~al.}}

\begin{frontmatter}

\title{Pseudo-healthy synthesis with pathology disentanglement and adversarial learning\tnoteref{tnote1}}%

\author[1]{Tian \snm{Xia}\corref{cor1}}
\cortext[cor1]{Corresponding author. }
\ead{tian.xia@ed.ac.uk}
\author[1]{Agisilaos \snm{Chartsias}}
\author[1,2]{Sotirios A. \snm{Tsaftaris}}

\address[1]{Institute for Digital Communications, School of Engineering, University of Edinburgh, West Mains Rd, Edinburgh EH9 3FB, UK}

\address[2]{The Alan Turing Institute, London, UK}


\begin{abstract}
Pseudo-healthy synthesis is the task of creating a subject-specific `healthy' image from a pathological one. Such images can be helpful in tasks such as anomaly detection and understanding changes induced by pathology and disease. In this paper, we present a model that is encouraged to disentangle the information of pathology from what seems to be healthy.
We disentangle what appears to be healthy and where disease is as a segmentation map, which are then recombined by a network to reconstruct the input disease image. We train our models adversarially using either \textit{paired} or \textit{unpaired} settings, where we pair disease images and maps when available. We quantitatively and subjectively, with a human study, evaluate the quality of pseudo-healthy images using several criteria. We show in a series of experiments, performed on ISLES, BraTS and Cam-CAN datasets, that our method is better than several baselines and methods from the literature. We also show that due to better training processes we could recover deformations, on surrounding tissue, caused by disease. Our implementation is publicly available at \url{https://github.com/xiat0616/pseudo-healthy-synthesis}. 


\end{abstract}


\end{frontmatter}


\section{Introduction}
\label{sec1}

Pseudo-healthy synthesis aims to  generate subject-specific `healthy' images from pathological ones.  By definition, a good pseudo-healthy image should both be \textit{healthy} and preserve the subject \textit{identity}, i.e. belong to the same subject as the input. The synthesis of such `healthy' images has many potential applications both in research and clinical practice. For instance, synthetic `healthy' images can be used for pathological segmentation, e.g. ischemic stroke lesion, by comparing the real with the synthetic image~\citep{ye2013modality,bowles2017brain}. Similarly, these `healthy' images can be used for detecting which part of the brain is mostly affected by neurodegenerative diseases, e.g. in Alzheimer disease, a more challenging task because of the global effect of these diseases~\citep{baumgartner2018visual}. 



However, devising methods that achieve the above task remains challenging.  Methods relying on supervised learning are not readily applicable, as finding both pathological and healthy images of the same subject for training and evaluation is not easy, since a subject cannot be `healthy' and `unhealthy' at the same time. Even though the use of longitudinal data could perhaps alleviate this, the time difference between observations would introduce more complexity to the task by adding as a confounder ageing alterations on the images beyond the manifestation of the actual disease. 

Prior to the rise of deep learning, approaches were focused on learning manifolds between `healthy' and `diseased' local regions at the patch \citep{ye2013modality,tsunoda2014pseudo} or even voxel level \citep{bowles2016pseudo}. However, the extent that these methods could capture global alterations of appearance, due to disease, remained limited.  

Recently though, the advent of deep learning in medical imaging \citep{litjens2017survey} has led to new approaches to pseudo-healthy synthesis. \citet{schlegl2017unsupervised} and \citet{chen2018unsupervised} for example, scaled up the approach of manifold learning to the image level with convolutional architectures.  More recently, adversarial approaches allowed learning mappings between the healthy and pathological image domains \citep{baumgartner2018visual,sun2018adversarial}.

\subsection{Motivation for our approach} \label{sec1_motivation}

\begin{figure}[t]
    \centering
    \includegraphics[scale=0.45]{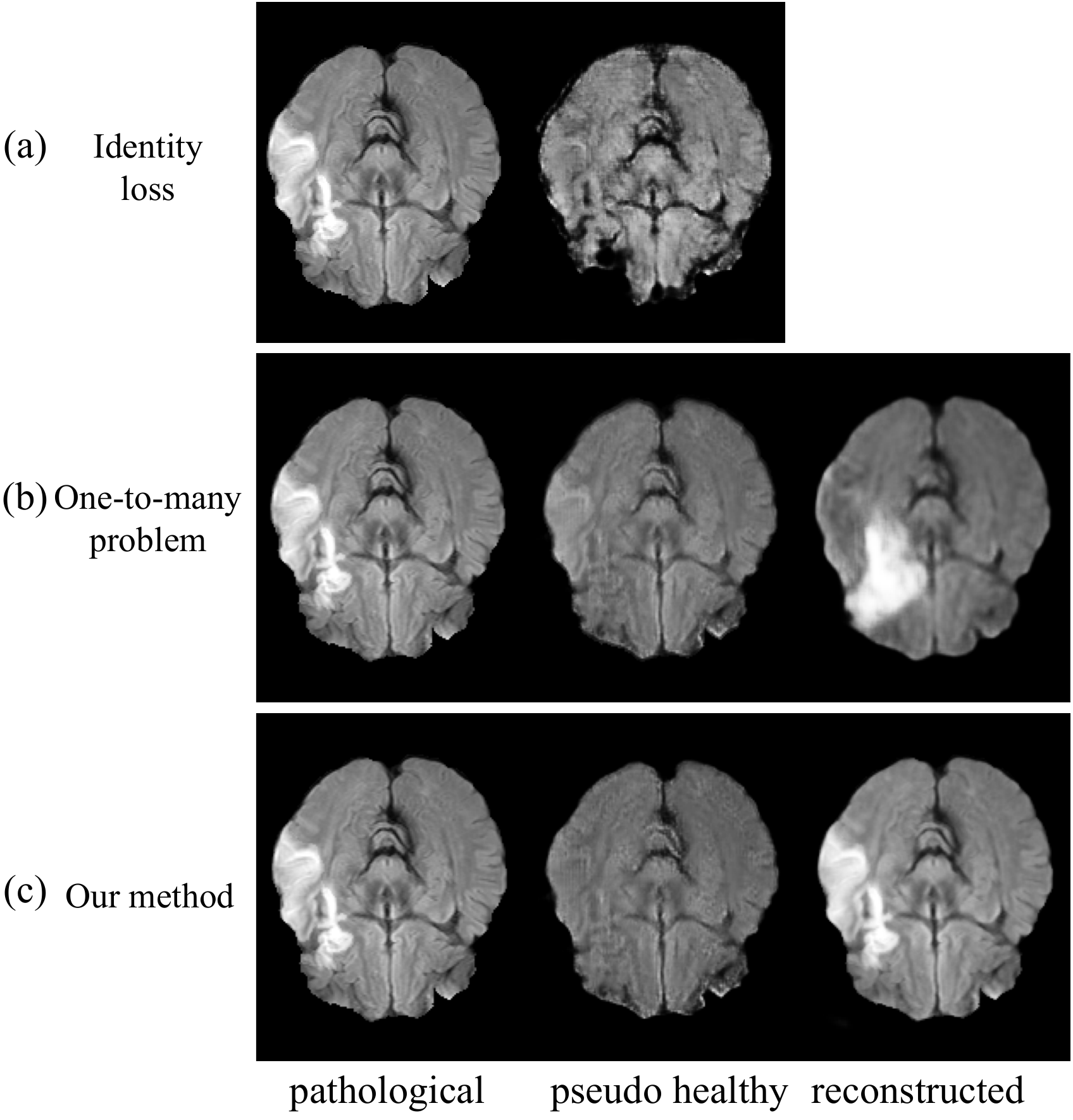}
    \caption{ The challenge of preserving identity. (a) shows an example of \textit{identity} loss in the generated `healthy' image. (b) shows a failure example of \textit{one-to-many problem} (described in Section \ref{sec:one_to_many_problem}). (c) shows an example obtained by our method which preserves \textit{identity} well. From left to right are the pathological image, pseudo-healthy image and the reconstructed image (if any), respectively. The example is taken from the ISLES dataset. }
    \label{fig:cyclegan_failure}
\end{figure}

We follow the same spirit, but differently from previous works our method focuses on disentangling the pathological from the healthy information, as a principled approach to guide the synthetic images to be `healthy' and preserve subject \textit{`identity'}.  Figure~\ref{fig:cyclegan_failure}(a) illustrates an example of identity loss. Thus, while our goal is to come up with an image that is healthy looking, we also aim to preserve identity such that the generated image belongs to the same input subject. 


We use cycle-consistency \citep{zhu2017unpaired} to help preserve identity but this introduces the so-called \textit{one-to-many problem} (detailed description in Section \ref{sec:one_to_many_problem}), where due to lack of information in the pseudo-healthy image we may now lose identity in the reconstructed image (see Figure \ref{fig:cyclegan_failure}(b)).  Our approach, by disentangling the information related to disease in a separate segmentation mask, circumvents this and helps enable many-to-many mappings (see Figure \ref{fig:cyclegan_failure}(c)). 


\subsection{Overview for our approach}

\begin{figure}[!t]
    \centering
    \includegraphics[scale=0.37]{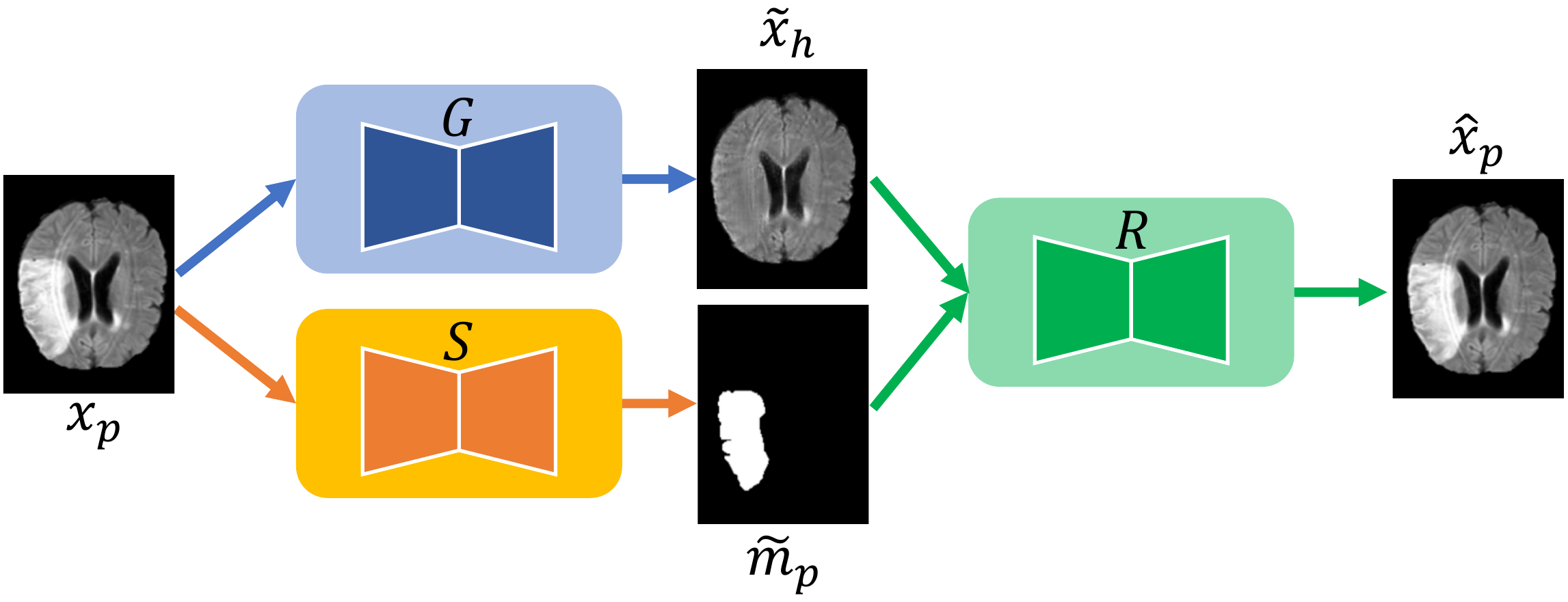}
    \caption{Schematic of our approach. A pseudo-healthy image $\tilde{x}_{h}$ is generated from the input pathological image $x_{p}$ by the \textit{Generator} (G); a pathological mask $\tilde{m}_{p}$ is segmented from $x_{p}$ by the \textit{Segmentor} (S); finally a reconstructed image $\hat{x}_{p}$ is reconstructed from  $\tilde{x}_{h}$ and $\tilde{m}_{p}$ by the Reconstructor (R). }
    \label{fig: schematic of method}
\end{figure}

A simple schematic of our proposed 2D method is shown in Figure \ref{fig: schematic of method}. The proposed network contains three components to achieve our goal during training: the \textit{Generator} (G) transforms a pathological image to a pseudo-healthy one; the \textit{Segmentor} (S) segments the pathology in the input image; finally, the \textit{Reconstructor} (R) reconstructs the input pathological image by combining the `healthy' image with the segmented mask and closes the cycle. The segmentation path is important to preserve the pathological information, and the reconstruction path involving the cycle-consistency loss contributes to the preservation of the subject identity. Note that during inference we only use the Generator and Segmentor.

The proposed method can be trained in a supervised manner using \textit{paired} pathological images and masks. However, since manually annotating pathology can be time-consuming and requires medical expertise, we also consider an \textit{unpaired} setting, where such pairs of images and masks are not available. Overall, our method is trained with several losses including a cycle-consistency loss \citep{zhu2017unpaired}, but we use a modified second cycle where we enforce healthy-to-healthy image translation to help preserve the identity.

\subsection{Contributions}
\label{sec: contribution}
The main contributions of this work are the following: 
\begin{itemize}
\item We propose a method for pseudo-healthy synthesis by disentangling anatomical and pathological information, with the use of supervised and unsupervised (adversarial)  costs.


\item Our method can be trained in two settings: \textit{paired} in which pairs of pathological images and masks are available, and \textit{unpaired} in which there are no corresponding segmentations for the input images.  


\item We introduce quantitative metrics\footnote{Most existing works on pseudo-healthy synthesis do not directly focus on the quality of the synthetic images but offer indirect evaluation: either through  performance improvements (if any) on  downstream tasks, or qualitatively with visual examples. Herein, since the application of pseudo-healthy synthesis heavily relies on the fidelity of the synthesised image, we directly evaluate it.} and subjective studies to evaluate the `healthiness' and `identity' of the synthetic results, and present extensive experiments comparing with four different methods (baselines and recent models form the literature), as well as ablation studies, on different MRI modalities.

\item We observe that our method may have the capacity of correcting brain deformations caused by high grade glioma, and propose a metric to assess this deformation correction.


\end{itemize}

In this paper, we advance our preliminary work \citep{xia2019adversarial} considerably: 1) we employ a different adversarial loss, namely Wasserstein GAN with gradient penalty \citep{gulrajani2017improved}, that improves image quality, offers more stable training and allows to correct for deformations due to the presence of disease; 2) we also use an additional `healthy' dataset, Cam-CAN, that improves training; 3) we offer more experiments and a detailed analysis of performance, including new metrics and two additional methods from the literature that we compared with; and 4) we introduce a subjective study where human raters evaluate the quality of created images.

The rest of the paper is organised as follows: Section \ref{sec2} reviews the literature related to pseudo-healthy synthesis. Section \ref{sec3} presents our proposed method. Section \ref{sec4} describes the experimental setup and Section \ref{sec: reults} presents the results and discussion. Finally, Section \ref{sec6} concludes the manuscript.

\section{Related work}
\label{sec2}

The concept of medical image synthesis is defined by \cite{special2018Frangi} as `\textit{the generation of visually realistic and quantitatively accurate images}', and the corresponding task has attracted significant attention recently. Here, we briefly review literature related to pseudo-healthy synthesis using non-deep learning (Section \ref{sec2_1}), but then turn our focus to deep learning techniques that learn a manifold of healthy data based on autoencoder formulations (Section \ref{sec2_2}). More related to our method, we review techniques that apply generative adversarial networks to pseudo-healthy synthesis (Section \ref{sec2_3}). Finally, we conclude this section with the differences between our method and these approaches (Section \ref{sec2_4}).

\subsection{Non-deep learning methods} \label{sec2_1}

Early methods learned local manifolds at the patch or pixel level. Patches were used together with dictionary learning to learn a linear mapping of source (pathological) and target (healthy) patches. Then, pseudo-healthy synthesis can be performed by searching for the closest patches within the dictionary and propagating the corresponding healthy patches to the synthetic `healthy' image. For example, \citet{ye2013modality} synthesised pseudo-healthy T2 images from T1 images.  Similarly, \citet{tsunoda2014pseudo} created a dictionary of normal lung patches and performed pseudo-healthy synthesis as a way to detect lung nodules. However, these methods heavily rely on the variation and size of the learned dictionaries. When input pathological patches are not similar to the training patches, these methods may not find suitable healthy patches to generate the `healthy' image. Furthermore, these methods are limited by the linear approximation of the dictionary decomposition.

Regression-based methods, instead, map intensities from one domain to another. A classical example is the method of \citet{bowles2017brain}, in which kernel regression maps T1-w images to FLAIR, exploiting the fact that pathology is not dominant in T1-w modality, in the domain tested. Note that this may not be true in all cases and not when translating to the same modality.

\subsection{Autoencoder methods} \label{sec2_2}

Aiming to scale up the receptive field of these methods and to permit more complex non-linear mappings, deep learning methods were employed first by learning compact manifolds in latent spaces to represent healthy data employing autoencoders   \citep{schlegl2017unsupervised,baur2018deep,uzunova2019unsupervised,you2019unsupervised, chen2018unsupervised}. These approaches assume that  when abnormal images are given to a neural network trained with healthy data, they are transformed (via the reconstruction function of the autoencoder) to images within the normal (healthy) distribution.
Usually non-healthy data are not used in training and guarantees that the synthetic images will maintain subject identity and be indeed within the manifold of the healthy distribution are thus not given. Furthermore, recently the correctness of modelling an input (normal) distribution to detect abnormal, out-of-distribution data has been questioned~\citep{nalisnick2019deep}.  

\subsection{Generative models} \label{sec2_3}

To involve abnormal data, Generative Adversarial Networks (GANs) \citep{goodfellow2014generative} and variants \citep{chen2016infogan, zhu2017unpaired} can be used. In its simplest form a Conditional GAN \citep{mirza2014conditional} can be used to translate pathological to healthy images without the need for input-output pairs (i.e.\ unpaired). However, since it focuses on synthesising an output within the target distribution, it may not guarantee the preservation of the subject's identity. 


To help encourage the preservation of identity some regularization is necessary. \citet{isola2017image} and \citet{baumgartner2018visual} used a $\ell_1$ regularization loss, along with an adversarial loss to help preserve identity. However, \citet{isola2017image} had access to paired training data, and thus applied the regularization loss to the output and target  images. Due to lack of paired data in the medical domain, \citet{baumgartner2018visual} minimised this regularization loss between input (pathological) and output images (pseudo-healthy). One potential problem with this could be that the regularization loss conflicts with the synthesis process. To offer an example, \citet{baumgartner2018visual} focused on the visual attribution of Alzheimer's Disease, where the disease effect is diffuse, and set a large weight for the regularization loss to ensure identity preservation. But in other cases (e.g.\ glioblastoma and ischemic stroke), where the disease effect can be significant (and perhaps localised and not diffused), it is difficult to balance the adversarial loss (which aims to change the input image to make it `healthy') and the regularization loss (which aims to minimise the change). If the emphasis on regularization is strong, then the network may not be able to make sufficient changes for accurate pseudo-healthy synthesis. On the contrary, if the weight of the regularization loss is small, then the identity might be compromised.

An approach to help preserve identity in the unpaired setting is the cycle-consistency loss of CycleGAN \citep{zhu2017unpaired}.
CycleGAN has been adopted for pseudo-healthy synthesis of glioblastoma brain images \citep{ cohen2018distribution, andermatt2018pathology, vorontsov2019boosting} and for liver tumours \citep{sun2018adversarial}. However, when one domain contains less information than the other, CycleGAN faces the \textit{one-to-many problem} (described in Section 3.2, which affects the quality of synthetic images, as mentioned in Section~\ref{sec1_motivation} and highlighted in Figure~\ref{fig:cyclegan_failure}(b). In order to alleviate this problem, \citet{andermatt2018pathology} and \citet{vorontsov2019boosting} provided pathology as residual and treated tumour as an additive factor. Specifically, when mapping healthy images to  pathological images, \citet{andermatt2018pathology} and \citet{vorontsov2019boosting} first randomly sample a pathological residual which is then added to the input healthy image to obtain the synthetically generated pathological image. Since this might prove difficult in fixing deformations, both papers mainly focused on achieving good segmentations, and not on the quality of the pseudo-healthy images. Our approach differs from these two methods by treating pathology as a complex factor that can affect the whole brain. In addition, part of the training process involves the Cycle H-H,  detailed in Section 3.5, to help synthesis.  

\subsection{Our approach} 
\label{sec2_4}
Our approach aims to address the above shortcomings. Similar to CycleGAN, our approach uses cycle-consistency losses to encourage identity preservation, however it also addresses the \textit{one-to-many problem} by disentangling images in pathological and anatomical factors. Thus, we aim to control both processes. In addition, in our effort to demonstrate the capabilities of adversarial approaches, we use as healthy domain images from a different unrelated dataset. This helps correct deformations caused by tumour masses. Finally, as we also noted in Section~\ref{sec1}, we directly evaluate images explicitly with new metrics, as well as with an observer study, rather than implicitly evaluating quality with performance in downstream tasks.

\section{Materials and methods} \label{sec3}

\begin{figure*}[t!]
    \centering
    \includegraphics[scale=0.5]{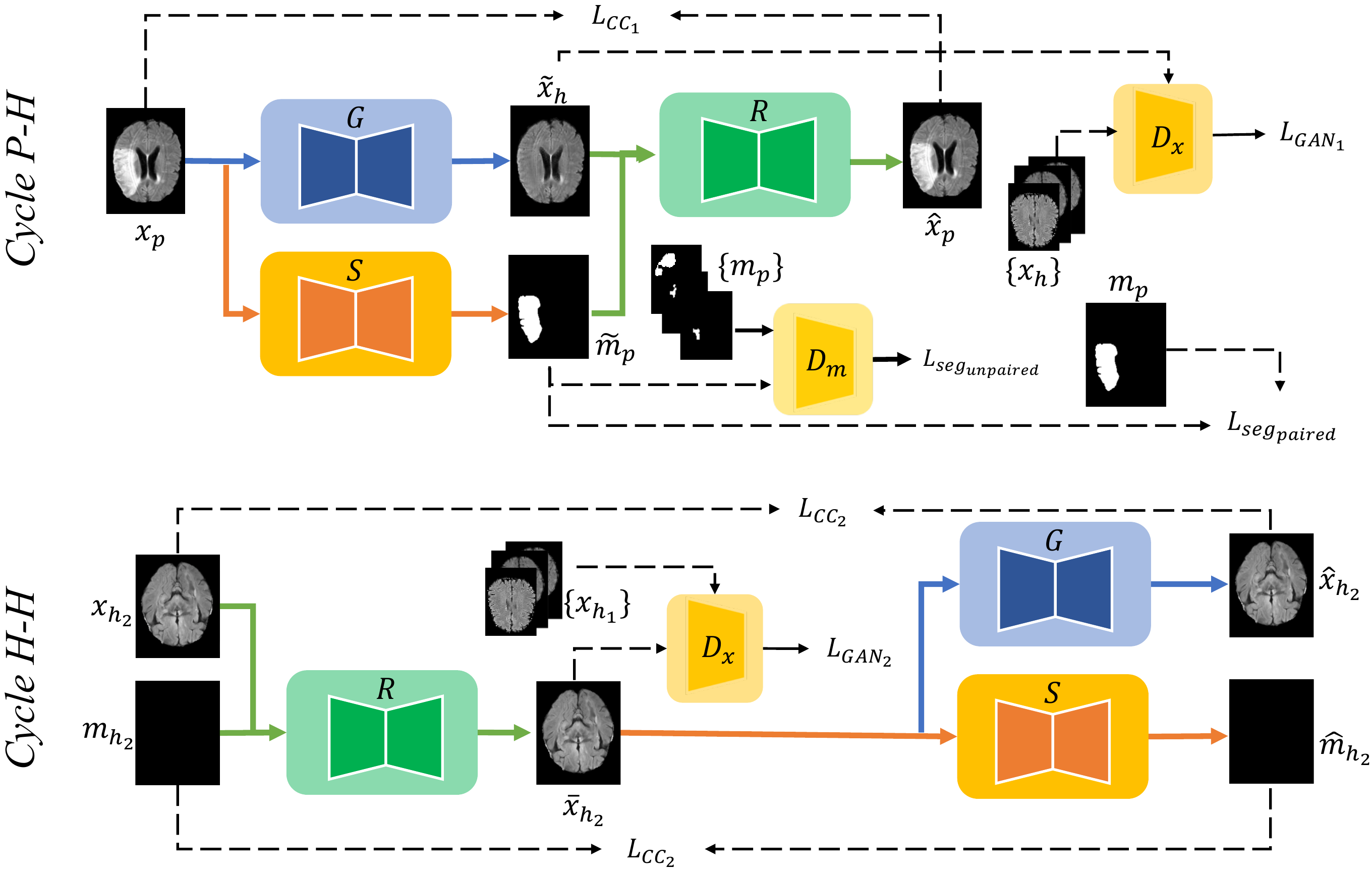}
    \caption{Training the proposed method. In \textit{Cycle P-H}, a pathological image $x_{p}$ is firstly disentangled into a corresponding pseudo-healthy image $\tilde{x}_{h}$ and a pathology segmentation $\tilde{m}_{p}$. Synthesis is performed by the generator network $G$ and the segmentation by the segmentor $S$. The pseudo-healthy image and the segmentation are further combined in the reconstructor network $R$ to reconstruct the pathological image $\hat{x}_{p}$. In Cycle H-H, a healthy image $x_{h}$ and its corresponding pathology map (a black mask) $m_{h}$ are put to the input of the reconstructor $R$ to get a fake `healthy' image, denoted as $\bar{x}_{h}$ to differ from the pseudo-healthy image $\tilde{x}_{h}$ in \textit{Cycle P-H}. This `healthy' image $\bar{x}_{h}$ is then provided to \textit{G} and \textit{S} to reconstruct the input image and mask, respectively.  }
    \label{fig:method_overview}
\end{figure*}

\subsection{Problem overview and notation}

We denote a pathological image as $x_{p_{i}}$, $i$ indicating a subject. $x_{p_{i}}$ belongs to the pathological distribution, $x_{p_{i}} \sim \mathcal{P}$. The goal is to generate a pseudo-healthy image $\tilde{x}_{h_{i}}$ for the pathological image $x_{p_{i}}$, such that $\tilde{x}_{h_{i}}$ lies in the distribution of healthy images, $\tilde{x}_{h_{i}} \sim \mathcal{H}$. We also want the generated image $\tilde{x}_{h_{i}}$ to maintain the identity of subject $i$. Therefore, pseudo-healthy synthesis can be formulated as two major objectives: \textit{remove} the disease of pathological images, and \textit{maintain} the identity and realism. For ease and unless explicitly stated, in the rest of the paper, we omit the subscript index $i$, and directly use $x_{p}$ and $x_{h}$ to represent samples from $\mathcal{P}$ and $\mathcal{H}$ distributions, respectively.


\subsection{The one-to-many problem: motivation for pathology disentanglement}
\label{sec:one_to_many_problem}

The transformation of a pathological image $x_{p}$ to its healthy version $\tilde{x}_h$ means that $\tilde{x}_h$ does not have the information of pathology present in the image. The question that arises is then: \textit{How can CycleGAN reconstruct $x_{p}$ from $\tilde{x}_h$ when this pathology information is lost?} There could be many $x_{p}$ with disease appearing in different locations that correspond to the same $\tilde{x}_h$. Given this information loss from one domain to the other, CycleGAN has to either hide information within the domain data \citep{chu2017cyclegan} and/or somehow within the extra capacity of the network to `permit' it to invent the missing information. An example failure case can be seen in Figure \ref{fig:cyclegan_failure}(b). We observe that the location and shape of the ischemic lesion is different between the original and reconstructed image. This is because the pseudo-healthy image does not contain, anymore, lesion information to guide the reconstruction of the input image.  

Recent papers \citep{chartsias2018b, almahairi2018augmented,chartsias2019disentangled}
have shown that auxiliary information can be provided in the form of a style or modality specific code (a vector) to guide the translation and permit now a well-posed one-to-one mapping. Our paper follows a similar idea and considers the auxiliary information to be spatial, and specifically stores the location and shape of the pathology in the form of a segmentation map. This then overcomes the one-to-many problem, and prevents the decoder from storing disease related features in the weights and the encoder from the need to encode pathology information in the pseudo-healthy image.

\subsection{Proposed approach}
\begin{figure*}[!t]
    \centering
    \includegraphics[scale=0.48]{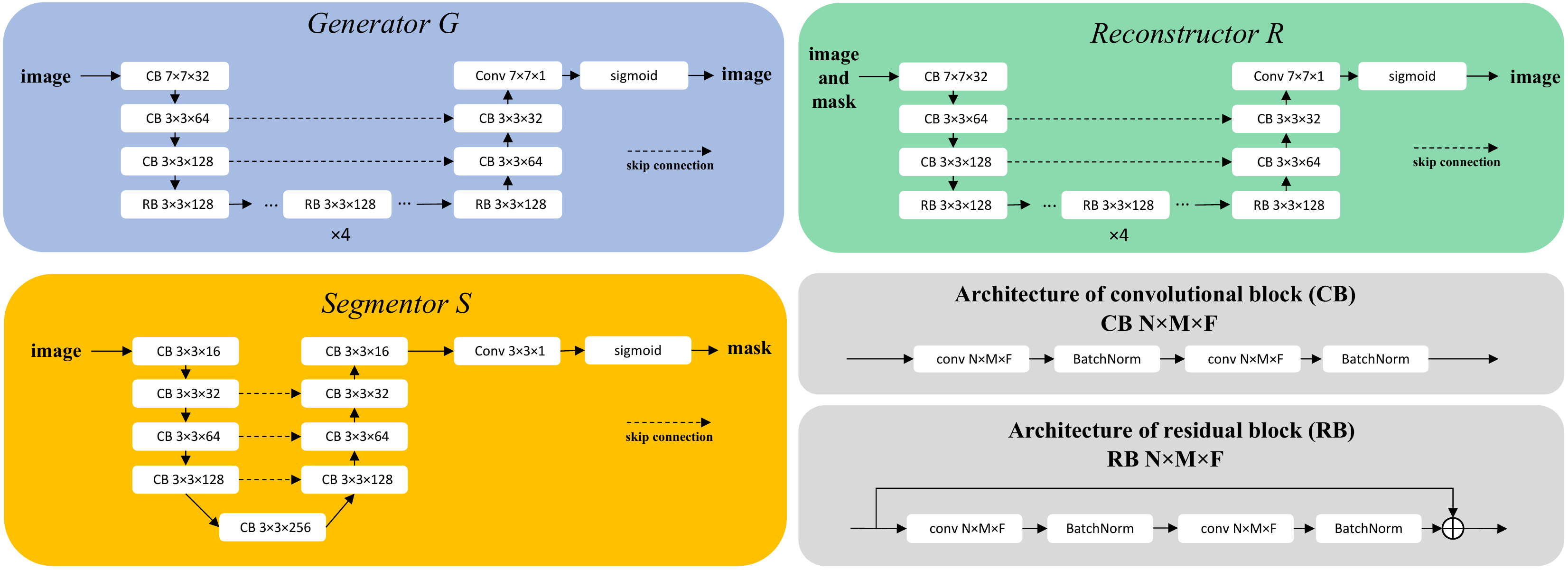}
    \caption{Detailed architectures of three main components in our method. The \textit{Generator G} and \textit{Reconstructor R} are modified residual networks \citep{he2016deep} with long skip connections between up- and down-sampling blocks. The difference between the Generator and the Reconstructor is that the first takes a one-channel input (image), whereas the second takes a two-channel input (image and mask). The Segmentor is a U-net \citep{ronneberger2015u} with long skip connections. All convolutional layers use \textit{LeakyReLU} as activation function, except for the last layers which use \textit{sigmoid}.}
    \label{fig: detailed structure of components}
\end{figure*}

An overview of our approach including the training losses is illustrated in Figure \ref{fig:method_overview}. 
The proposed method contains three components, the architectures of which are shown in Figure \ref{fig: detailed structure of components}: the \textit{Generator}, the \textit{Segmentor} (S) and the \textit{Reconstructor} (R).  The Generator and the Segmentor comprise the pseudo-healthy part of our approach, and disentangle a diseased image into its two components, the corresponding pseudo-healthy image and the segmentation mask. 

%

\subsubsection{Generator}
The Generator transforms diseased to pseudo-healthy images. Differently from our previous work \citep{xia2019adversarial}, which used a residual network \citep{he2016deep} with downsampling and upsampling paths, the new Generator architecture has long skip connections between downsampling and unsampling blocks. This helps better preserve details of the input images and results in sharper outputs. The detailed architecture of the Generator is shown in Figure \ref{fig: detailed structure of components}.

\subsubsection{Segmentor}
The Segmentor  predicts a binary disease segmentation map.\footnote{We also investigated using a single neural network with shared layers and two outputs  to perform this decomposition, but found that using two separate networks enables more stable training. This architectural choice is in line with other disentanglement methods \citep{huang2018munit, lee2018diverse}.
} This map helps localise and delineate disease in the
reconstructed image. The Segmentor follows a U-net \citep{ronneberger2015u} architecture, shown in Figure \ref{fig: detailed structure of components}.

\subsubsection{Reconstructor}
The Reconstructor takes a pseudo-healthy image and a corresponding segmentation mask of the disease, concatenates them in a two-channel image, and reconstructs the input, pathological, image. The architecture of the Reconstructor is the same as the one of the Generator, except that Generator takes one-channel input but Reconstructor takes a two-channel input. Image reconstruction is key for our method since it encourages the preservation of subject identity.

\subsubsection{Discriminators}
Our method involves two discriminators that are used in adversarial training. One is the discriminator for pseudo-healthy images (denoted as $D_{x}$) which  encourages generation of realistic pseudo-healthy images. The other is used to help learn a manifold for the pathology mask (denoted as $D_{m}$) which is used to train the Segmentor when paired pathological images and masks are not available (more details in Section \ref{sec: paired and unpaired}). The architecture of both discriminators follow the design used by \citet{baumgartner2018visual}. The adversarial training is performed with a Wasserstein loss with gradient penalty \citep{gulrajani2017improved}. 


\subsection{Model training}
Inspired by \citet{zhu2017unpaired}, we involve two cycles to train our model, which are shown in Figure \ref{fig:method_overview}.  The first cycle is \textit{Cycle P-H}, where we perform pseudo-healthy synthesis. The Generator G first takes a pathological image $x_{p}$ as input, and produces a pseudo-healthy image: $\tilde{x}_{h}=G(x_p)$. Similarly, the Segmentor S takes $x_{p}$ as input and outputs a mask $\tilde{m}_{p}$ indicating where the pathology is: $\tilde{m}_{p}=S(x_p)$. The Reconstructor R then takes both $\tilde{x}_{h}$ and $\tilde{m}_{p}$ as input and generates a reconstruction of the input image: $\tilde{x}_{p}=R(\tilde{x}_{h}, \tilde{m}_{p})$.

The second cycle is \textit{Cycle H-H} which is designed to stabilise training, help preserve input identity, and further encourage disentanglement of disease from the pseudo-healthy image. The Reconstructor first takes as input a healthy image $x_{h}$ and a `healthy' mask $m_{h}$ ,i.e.\ an image of all zeros, and produces a fake healthy image: $\bar{x}_{h}=R(x_{h}, m_{h})$. This fake healthy image $\bar{x}_{h}$ is then passed as input to the Generator G, $\hat{x}_{h}=G(\bar{x}_{h})$, and Segmentor to reconstruct the input healthy image and mask, $\hat{m_h}=S(\bar{x}_{h})$, respectively.

The design of Cycle H-H is due to several reasons. 
First, we want to ensure that the Reconstructor does not invent pathology when given a healthy mask as input. Second, we encourage the Generator to better preserve identity, i.e. when the input to G is a `healthy' image, the output should be the same `healthy' image. Similarly, when given a `healthy' image, the Segmentor should not detect any pathology. When the predicted output is not a black map, it means that either the Reconstructor is not trained well, i.e. it creates pathology-like artefacts, or the Segmentor is not trained well, i.e. it finds non-existing pathology. In this case, the Reconstructor and Segmentor are penalised. This in turn also encourages the Segmentor not to hide  information  useful for reconstruction, and thus any anatomical information is only contained in the pseudo-healthy image.\footnote{We note here that we could also have considered a cycle where we could take a pseudo-healthy image and pass it through the segmentor and penalise if any disease pixels are detected. We found that this is less stable: either the segmentor could have thrown a false positive or the generator made an error. We found the design of the current Cycle H-H more robust and our experiments show that the pseudo-healthy images rarely contain detectable, by a judge segmentor, disease pixels.}

\subsection{Paired and unpaired settings}
\label{sec: paired and unpaired}
There are two settings of training the Segmentor (S) considering the availability of ground-truth pathology labels.

In the first, termed \textit{paired} setting, we have paired pathological images and ground-truth masks. In this setting, we train the Segmentor  directly using the ground-truth pathology masks with a differential analogue of the Dice segmentation loss. 

In the second, termed \textit{unpaired} setting, we do not have pairs of pathological images and masks. In this setting, since supervised training is not feasible, we involve a \textit{Mask Discriminator} termed as $D_{m}$ that distinguishes segmented masks from real pathology masks, and thus learns a prior on the pathology shape. The Segmentor is then trained adversarially against this Mask Discriminator. The real pathology masks used for training are ground-truth pathology masks chosen randomly from other subjects. The losses are described mathematically for each setting in Section \ref{sec: segmentation losses}.

\subsection{Losses}
The training losses can be divided into three categories, \textit{adversarial losses}, \textit{cycle-consistency losses} and \textit{segmentation losses}, the details of which are described below.

\subsubsection{Adversarial losses for images}

The synthesis of pseudo-healthy image $\tilde{x_{h}}$ ($\tilde{x_{h}}=G(x_{p})$) in \textit{Cycle P-H} is trained using the Wasserstein loss with gradient penalty \citep{gulrajani2017improved}:
\begin{equation}
\label{eq: L_gan_1}
\begin{split}
L_{GAN\textsubscript{1}}=
\stackunder{max}{\textit{D\textsubscript{x}}}\, \stackunder{min}{\textit{G}}\, \E_{x_{p}\sim  \mathcal{P}, \, x_{h} \sim \mathcal{H}}[D_{x}(x_{h})-D_{x}(G(x_{p}))\\+ \lambda_{GP}({\lVert}\nabla_{\dot{x}_{h}}(\dot{x}_{h}){\rVert}_{2}-1)^2],
\end{split}
\end{equation}
where $x_{p}$ is a pathological image, $G(x_{p})$ is its corresponding pseudo-healthy image, $x_{h}$ is a healthy image, $D_{x}$ is the discriminator to separate real and fake samples, and $\dot{x}_{h}$ is the average sample defined by $\dot{x}_{h}=\epsilon \, x_{h} + (1-\epsilon)\, G(x_{p})$,  $\epsilon \sim  U[0,1]$. The first two terms measure the Wasserstein distance between real healthy and synthetic healthy images; the last term is the gradient penalty loss involved to stabilise training.  As in \citet{gulrajani2017improved} and \citet{baumgartner2018visual}, we set $\lambda_{GP}=10$.

Similarly, we have $L_{GAN_{2}}$ for the fake `healthy' image $\bar{x}_{h}$ ($\bar{x}_{h}=R(x_{h},m_{h})$) in Cycle H-H:
\begin{equation}
\label{eq: L_gan_2}
\begin{split}
L_{GAN\textsubscript{2}}=\stackunder{max}{\textit{D\textsubscript{x}}}\, \stackunder{min}{\textit{R}}\, \E_{x_{h_{1}}\sim \mathcal{H},\, 
x_{h_{2}}\sim \mathcal{H},\, m_{h_2} \sim \mathcal{H}_m}[D_{x}(x_{h_{1}})\\ -D_{x}(R(x_{h_{2}}, m_{h_{2}}))+ \lambda_{GP}({\lVert}\nabla_{\dot{x}_{h}}(\dot{x}_{h}){\rVert}_{2}-1)^2],
\end{split}
\end{equation}
 where $x_{h_{1}}$ and $x_{h_{2}}$ are two different healthy images drawn from the healthy image distribution $\mathcal{H}$, $m_{h_{2}}$ is the corresponding pathology mask of $x_{h_{2}}$, i.e. a black mask,   $R(x_{h_{2}}, m_{h_{2}})$ is the fake `healthy' image reconstructed with $x_{h_{2}}$, and  $\dot{x}_{h}$ is defined as $\dot{x}_{h}=\epsilon \, x_{h_{1}} + (1-\epsilon)\, R(x_{h_{2}}, m_{h_{2}})$, $\epsilon \sim  U[0,1]$.
 
\subsubsection{Cycle-consistency losses}

 We involve cycle-consistency losses to help preserve the subject identity of the input images. For \textit{Cycle P-H}, we have:
 \begin{equation}
 \begin{split}
 \label{eq: loss_cc1}
     L_{CC_{1}} =\stackunder{min}{\textit{G,R,S}} \E_{x_{p}\sim \mathcal{P}}[{\lVert}R(G(x_{p}), S(x_{p}))-x_{p}{\rVert}_{1}],
 \end{split}
 \end{equation}
where $x_{p}$ is a pathological image, $G(x_{p})$ is the pseudo-healthy image produced by Generator, $S(x_{p})$ is the segmented pathology mask by Segmentor, $R(G(x_{p}), S(x_{p}))$ is the reconstructed pathological image by Reconstructor given $G(x_{p})$ and $S(x_{p})$. Similarly with \citet{zhu2017unpaired}, we use $\ell_1$ loss rather than $\ell_2$, to reduce the amount of blurring.

Similarly, for Cycle H-H, we have:

\begin{equation}
\begin{split}
    L_{CC_{2}} = \stackunder{min}{\textit{G,R,S}} \E_{x_{h_{2}}\sim \mathcal{H}, \, m_{h_{2}}\sim \mathcal{H}_m}[{\lVert} G(R(x_{h_{2}}, m_{h_{2}}))-x_{h_{2}}{\rVert}_{1}\\+{\lVert}S(R(x_{h_{2}}, m_{h_{2}}))-m_{h_{2}}{\rVert}_{1}],
\end{split}
\end{equation}
\noindent where $x_{h_{2}}$ and $m_{h_{2}}$ are a healthy image and the corresponding mask, respectively, $R(x_{h_{2}},m_{h_{2}})$ is the fake `healthy' image obtained by Reconstructor given a healthy image $x_{h_{2}}$ and a healthy mask $m_{h_{2}}$ as input, $G(R(x_{h_{2}}, m_{h_{2}}))$ is the reconstructed image by Generator given $R(x_{h}, m_{h_{2}})$, and  $S(R(x_{h_{2}}, m_{h_{2}}))$ is the segmented mask that corresponds to $R(x_{h_{2}},m_{h_{2}})$. Here we use $\ell_1$ loss for the reconstructed mask instead of the Dice loss as it is not well defined when the target masks are all black.

\subsubsection{Segmentation losses}
\label{sec: segmentation losses}
As described in Section \ref{sec: paired and unpaired}, there are two training settings for the Segmentor. For the paired setting where we have access to paired pathological image and masks, we use a supervised loss to train the Segmentor: 
\begin{equation}
\begin{split}
\label{eq: seg_paired}
    L_{seg_{paired}} = \stackunder{min}{\textit{S}}\E_{x_{p}\sim \mathcal{P},\, m_{p}\sim \mathcal{P}_{m}}[Dice( m_{p}, S(x_{p}))],
\end{split}    
\end{equation}
where $x_{p}$ and $m_{p}$ are paired pathological images and masks, $S(x_{p})$ is the predicted mask by \textit{Segmentor S}, and $Dice(.)$ represent the dice coefficient loss \citep{milletari2016v}.


In the unpaired setting, there are no paired  images and masks, and we use an adversarial loss to train the Segmentor:
\begin{equation}
\begin{split}
\label{eq: seg_unpaired}
    L_{seg_{unpaired}} = \stackunder{max}{\textit{D\textsubscript{m}}}\, \stackunder{min}{\textit{S}} \E_{x_{p_{1}}\sim \mathcal{P}, \, m_{p_{2}}\sim \mathcal{P}_{m} }[D_{m}(S(x_{p_{1}}))-D_{m}(m_{p_{2}})\\+\lambda_{GP}({\lVert} \nabla_{\bar{m}_{p}} D(\bar{m}_p){\rVert}_{2}-1 )^{2}],
\end{split}
\end{equation}
where $x_{p_{1}}$ is a pathological image, $m_{p_{2}}$ is a pathological mask randomly drawn from subjects other than $x_{p_{1}}$, $D_{m}$ is the discriminator to classify between the segmented mask $S(x_{p_{1}})$ and the randomly chosen mask $m_{p_{2}}$, and $\bar{m}_{p}$ is the average sample defined by  $\bar{m}_{p}=\epsilon \, m_{p_{2}} + (1-\epsilon)\, S(x_{p_{1}})$, $\epsilon \sim  U[0,1]$.

\section{Experimental setup}
\label{sec4}

\subsection{Data and pre-processing}
\label{sec: data and preprocessing}
\noindent \textbf{Data:} In this work we demonstrate our method on 2D slices from three datasets:
\begin{itemize}
    \item  \textit{Ischemic Stroke Lesion Segmentation  challenge 2015} contains 28 volumes which have been skull-stripped and resampled in an isotropic spacing of \textit{1 mm}, and co-registered to the FLAIR modality. All volumes have lesion segmentation annotated by experts. We use T2 and FLAIR modality for our experiment. 
    
    \item \textit{Multimodal Brain Tumor Segmentation Challenge 2018} (BraTS) \citep{menze2014multimodal} dataset contains high and low grade glioma cases. The tumour areas have been manually labelled by experts. All data have been skull-stripped, co-registered and resampled to \textit{1 mm} resolution.  In this work we select 150 volumes which contain high grade glioma/glioblastoma (HGG). The `healthy' slices in BraTS may not be really healthy, since the glioblastoma may affect areas of brain where it is not present \citep{menze2014multimodal}, for an example see Figure \ref{fig: brats healthy deformation}. We therefore involve Cam-CAN dataset as a healthy dataset, as described below. 
    
    \item \textit{Cambridge Centre for Ageing and Neuroscience} (Cam-CAN) \citep{taylor2017cambridge} dataset contains normal volumes from 17 to 85 years old. We randomly selected 76 volumes for our experiment. We chose to involve this dataset as `healthy' data when performing pseudo-healthy synthesis to avoid the possible deformations of brain tissues in BraTS images.  Since Cam-CAN only contains T1 and T2 modalities, we also use T1 and T2 from BraTS. 
    
\end{itemize}

\noindent \textbf{Pre-processing:} Initially, we skull-stripped the Cam-CAN volumes using FSL-BET \citep{jenkinson2005bet2}. We then linearly registered the Cam-CAN and BraTS volumes to MNI 152 space using FSL-FLIRT \citep{jenkinson2012fsl}.

We normalised the volumes of all datasets by clipping the intensities to $[0,V_{99.5}]$, where $V_{99.5}$ is the $99.5\%$ largest intensity value in the corresponding volume, and rescaled to the range $[0,1]$. We then selected the middle 60 axial slices from each volume, and cropped each slice to the size $[208,160]$.
For ISLES, we label a slice as `healthy' if its corresponding lesion map is black, otherwise as `pathological'. We label all slices from Cam-CAN as `healthy', and label a slice from BraTS as `pathological' if its corresponding pathology annotation is not a black mask, i.e. the glioblastoma is present in this slice. 

\noindent \textbf{Histogram check}:  We checked the histogram similarity between BraTS and Cam-CAN. Specifically, we normalised each histogram to a probability density distribution (PDF), and computed the Jensen\textendash Shannon (JS) divergence \citep{lin1991divergence} between the PDFs of the two datasets. We calculated a JS divergence of 0.009 between BraTS `healthy' slices (slices with no segmentations) and Cam-CAN slices, 0.011 between BraTS `healthy' and BraTS `pathological' slices, and 0.015 between BraTS `pathological' and Cam-CAN slices. This implies that after pre-processing, the difference between histograms of Cam-CAN and BraTS is minimal.





\subsection{Baselines and methods for comparison} \label{sec:baselines}
We compare our method with the following four approaches:
\begin{enumerate}
    \item \textbf{Conditional GAN:} We first consider a baseline that uses adversarial training and a simple conditional approach of \citet{mirza2014conditional}. This is a GAN in which the output is conditioned on the input image and does not use segmentation masks.  This baseline uses a generator and a discriminator with the same architectures as our method for appropriate comparison.  
    
    \item \textbf{CycleGAN:} Another baseline we compare with is the CycleGAN \citep{zhu2017unpaired}, where there are two translation cycles: one is \textit{P} to \textit{H} to \textit{P}, and the other is \textit{H} to \textit{P} to \textit{H} (`\textit{P}' refers to the pathological and `\textit{H}' refers to the healthy domain).   We do not use segmentation masks. 
    The generators and discriminators of CycleGAN also share the same architecture as our proposed method.  

    \item \textbf{AAE:} We implement and compare with a recent method that aims to address a similar problem  \citep{chen2018unsupervised}. We trained an adversarial autoencoder (AAE) only on healthy images and performed pseudo-healthy synthesis with the trained model.  This approach does not use segmentation masks and data with pathology. 

    \item \textbf{vaGAN:} We compare with \citet{baumgartner2018visual}, another recent method for pseudo-healthy synthesis, using the official implementation\footnote{\url{https://github.com/baumgach/vagan-code}} but modified for 2D slices.  This method produces residual maps, which are then added to the input images to produce the resulting pseudo-healthy images. An $\ell_2$ loss on the produced maps acts as a regulariser.  This approach does not use segmentation masks.

\end{enumerate}

\subsection{Training details}

In the paired setting, the overall loss is:
\begin{equation}
\begin{split}
    L_{paired}= \lambda_{1} L_{GAN_{1}}+\lambda_{2} L_{GAN_{2}}\\+ \lambda_{3} L_{CC_{1}} +\lambda_{4} L_{CC_{2}} +\lambda_{5} L_{seg_{paired}},
\end{split}  
\label{eq: losses in paired}
\end{equation}
where the $\lambda$ parameters are set to: $\lambda_{1}=2$, $\lambda_{2}=1$, $\lambda_{3}=20$, $\lambda_{4}=10$ and $\lambda_{5}=10$. 
In the unpaired setting, the loss is:
\begin{equation}
\begin{split}
    L_{unpaired}= \lambda_{1} L_{GAN_{1}}+\lambda_{2} L_{GAN_{2}}\\+\lambda_{3} L_{CC_{1}} +\lambda_{4} L_{CC_{2}}
    +\lambda_{5} L_{seg_{unpaired}},
\end{split}    
\label{eq: losses in unpaired}
\end{equation}
where $\lambda_{1}$, $\lambda_{2}$, $\lambda_{3}$ and $\lambda_{4}$ are set as above, while $\lambda_{5}$ is set to 1. The values of the $\lambda$ parameters are set experimentally and  similar to our previous work~\citep{xia2019adversarial} as follows. The $\lambda$ for Cycle P-H are double the $\lambda$ for Cycle H-H, i.e. $\lambda_1=2\lambda_2$ and $\lambda_3=2\lambda_4$, since our focus is on pseudo-healthy synthesis. Furthermore, the $\lambda$ for $L_{CC}$ is 10 times larger than the one for $L_{GAN}$ to balance the loss values, i.e. $\lambda_3=10\lambda_1$ and $\lambda_4=10\lambda_2$. Finally, $\lambda_5$ in paired setting is set to 10 to encourage an accurate segmentation, since segmentation is a challenging task. The $\lambda$ values for the unpaired setting are set similarly, except $\lambda_5$ that is set to 1, since this is a GAN loss, and a balance between the segmentor and mask discriminator losses is sought.

We train all models for 300 epochs. Following \citet{goodfellow2014generative} and \citet{arjovsky2017wasserstein}, we updated the discriminators and generators in an alternating session. As Wasserstein GAN requires the discriminators to be close to optimal during training, we updated the discriminators for 5 iterations for every generator update. Initially in the first 20 epochs, we update the discriminators for 50 iterations per generator update. We implemented our methods using Keras \citep{chollet2015keras}. We trained using \textit{Adam} optimiser \citep{kingma2015adam} with a learning rate of 0.0001 and $\beta_1$ equal to 0.5.  We will make our implementation publicly available at \url{https://github.com/xiat0616/pseudo-healthy-synthesis}.

The results of Section \ref{sec: reults} are obtained from a 3-fold cross validation. For ISLES, each split contains 18 volumes for training, 3 volumes for validation and 7 volumes for testing. For BraTS, each split contains 100 volumes for training, 15 for validation and 35 for testing. For Cam-CAN, each split contains 50 volumes for training, 8 for validation and 18 for testing. This is to ensure that the `pathological' slices from BraTS have similar number as the `healthy' slices from Cam-CAN.  We fine-tuned the architecture of the pre-trained segmentor and classifier based on the validation set.


\subsection{Evaluation metrics} 
\label{sec:evaluation}
Since paired healthy and pathological images of the same subjects are difficult to acquire, we do not have ground-truth images to directly evaluate the synthetic outputs. 


As we mentioned previously in Section~\ref{sec: contribution},  image quality has been rarely directly evaluated. To address this, previously, we proposed two numerical evaluation metrics to assess the `healthiness' and `identity' of synthetic images \citep{xia2019adversarial}. In this work, to evaluate how well the deformations are corrected in BraTS, we further propose a new metric and also perform a human evaluation study on a subset of our experiments. Below we introduce the new metric but for completeness we also (re)present healthiness and identity. 

\noindent \textbf{Healthiness ($h$):} To evaluate how `healthy' the pseudo-healthy images are, we measure the size of their segmented pathology as a proxy. To this end, we pre-trained a segmentor to estimate pathology from images. We then used this segmentor as a judge to assess pathology from the pseudo-healthy images and checked how large the estimated pathology areas are.  Note that for each split we trained a segmentor on the training data and fine-tuned it on the validation set.  Formally, \textit{healthiness} is defined as:
\begin{equation}
\label{eq: definition of healthiness}
\begin{split}
    \textit{h}=1-\frac{\E_{\hat{x}_h\sim\mathcal{H}}[N(f_{pre}(\hat{x}_{h}))]}{\E_{m_p\sim\mathcal{P}_m}[N(f_{pre}(x_{p}))]} =1- \frac{\E_{x_p\sim\mathcal{P}}[N(f_{pre}(G(x_{p})))]}{\E_{m_p\sim\mathcal{P}_m}[N(f_{pre}(x_{p}))]},
\end{split}    
\end{equation}
where $x_{p}$ is a pathological image, $f_{pre}$ is the pre-trained segmentor, and $N(.)$ is the number of pixels that are labelled as pathology by $f_{pre}$. The denominator uses the segmented mask of the pathological image $f_{pred}(x_{p})$, instead of the ground truth $m_p$, to cancel out a potential bias introduced by the pre-trained segmentor. We subtract the term from 1, such that when pathology mask gets smaller, \textit{h} increases.

\noindent \textbf{Identity ($iD$):} This metric represents how well the synthetic images preserve subject identity, i.e. how likely they come from the same subjects as the input images. This is achieved by evaluating their structural similarity to the input images outside the pathology regions, using a masked \textit{Multi-Scale Structural Similarity Index} (MS-SSIM)\footnote{Due to its use of MS-SSIM this metric also reflects image quality.} with window width of 11 \citep{wang2003multiscale}. 
Formally, \textit{identity} is defined as:
\begin{equation}
\begin{split}
    iD &= MS\text{-}SSIM[(1-m_{p})\odot \tilde{x_{h}}, (1-m_{p})\odot x_{p}] \\
    &= MS\text{-}SSIM[(1-m_{p})\odot G(x_{p}), (1-m_{p})\odot x_{p}],
\end{split}
\end{equation}
where $x_{p}$ is a pathological image, $m_{p}$ is its corresponding pathology mask, and $\odot$ is pixel-by-pixel multiplication. 

\noindent \textbf{Deformation correction ($DeC$):} 
In some cases  (BraTS dataset), a brain may also deform due to the presence of a large cancerous mass.  The difficulty is that, to fix the deformation caused by tumour, we need to not only change the abnormal intensities, but also to make necessary changes to the structure of the brain. This poses a significant challenge to measure the subject identity. The identity metric above does not measure well whether this tissue has recovered (because it relies on pixel correspondence). Herein we attempt to define a proxy metric that aims to assess whether such correction has taken place.\footnote{We note that this is a very hard task and our attempts to use a non-linear registration-based approach where we measured the amount of deformation between different diseased and pseudo-healthy images was not met with success because it gave lots of false positives when identity was completely lost.}

\begin{figure}[t]
    \centering
    \includegraphics[scale=0.4]{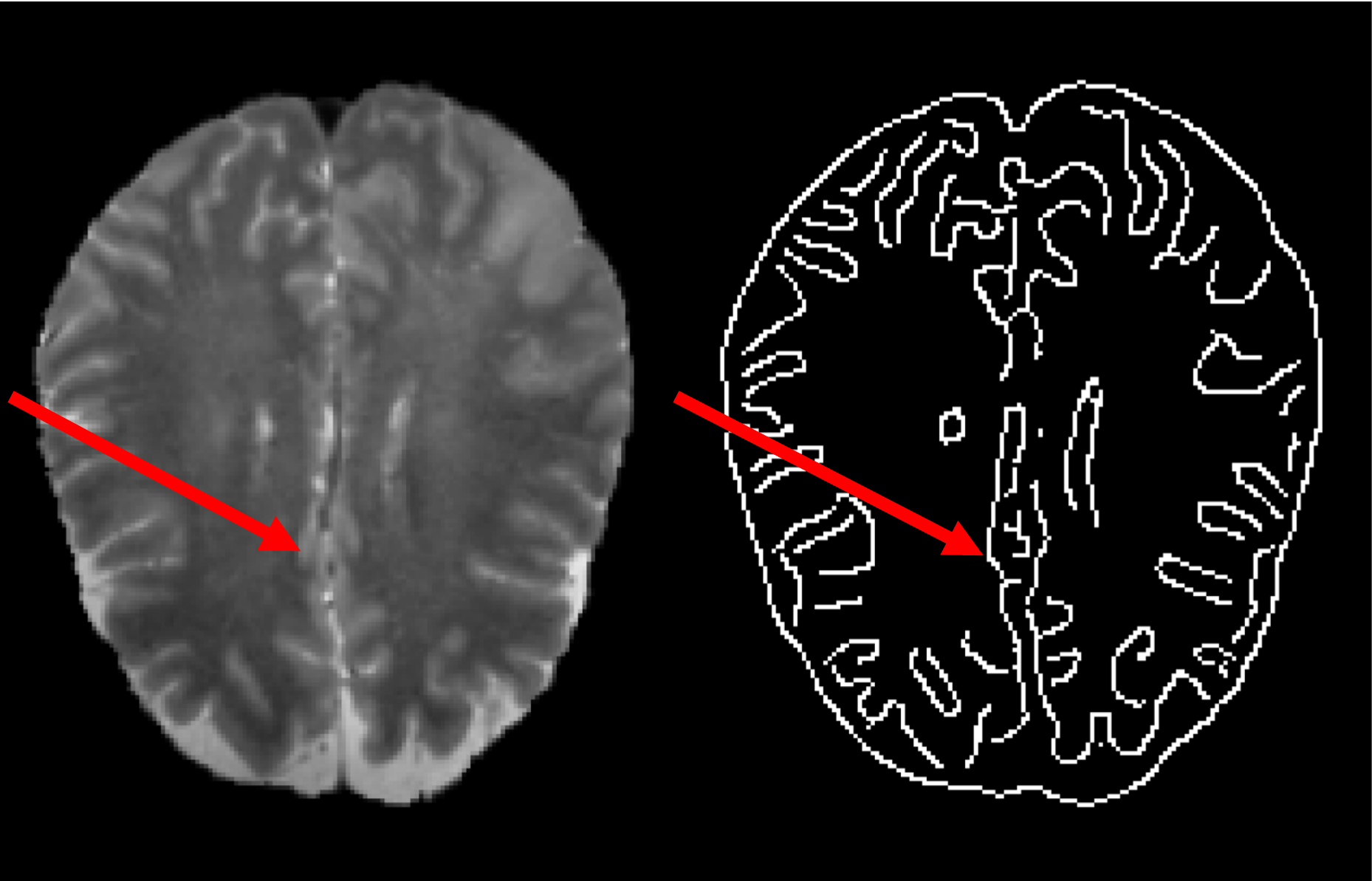}
    \caption{An example of BraTS `healthy' image and its edge map. Observe the deformation in the brain and edge as pointed out by the red arrows. Note that this brain image does not have pathology in its corresponding segmentation map, but the deformation still exists.}
    \label{fig:example_edge}
\end{figure}

As Cam-CAN and BraTS were acquired differently, and could potentially have intensity differences, we pre-processed all brain slices using the Canny edge detector in order to remove any intensity bias. An example of a BraTS image and its extracted edge map are shown in Figure~\ref{fig:example_edge}, where we can observe the deformations as pointed out by the red arrows. We then pre-trained a classifier to classify edge maps of BraTS `healthy' slices, i.e. images with no tumour annotation, and Cam-CAN slices. The pre-trained classifiers, achieved an average accuracy of 89.7\%, and were used as a judge on pseudo-healthy images from BraTS slices. This means that the classifiers were able to discriminate between BraTS `healthy' edges and Cam-CAN edges mostly relying on the presence of deformations.  The output of this classifier is a continuous number between 0 and 1, representing the probability of an image to be deformation-free. $DeC$ in the testing set is then defined as the probability  of synthetic images being deformation-free.

\noindent \textbf{Human evaluation:} To highlight the difficulty of defining quantitative metrics, and the overall difficulty of assessing image `quality' in such synthesis tasks, we introduce an expert evaluation to further assess the above criteria of healthiness, identity and deformation correction on a small subset of the experiments. We purposely did not ask raters to assess overall image quality, as quality can be a combination of factors (which can vary across experts).\footnote{We also note the difference of our study design compared to the ones commonly encountered in the image-to-image translation community \citep{zhu2017unpaired} where users are asked to decide if an image is `real' or `fake'.}

We randomly selected 50 slices from BraTS, obtained the pseudo-healthy outputs of all comparison methods, and then asked four medical image analysis researchers and a clinical neurologist  to independently score each synthetic image arranged in panels (details below) on each criterion using a binary score. We provided instructions as to what each criterion should reflect. Specifically the definitions were: ``Healthiness: assess if the synthetic image appears healthy (1) or not (0)''; ``Identity:  assess if the synthetic image belongs to the same subject as the original image (1) or not (0)''; ``Deformation correction: assess if the deformation caused by a cancerous mass has been corrected in areas outside the mass (1) or not (0)''. 
 
\begin{figure*}[b!]
    \centering
    \includegraphics[scale=0.8]{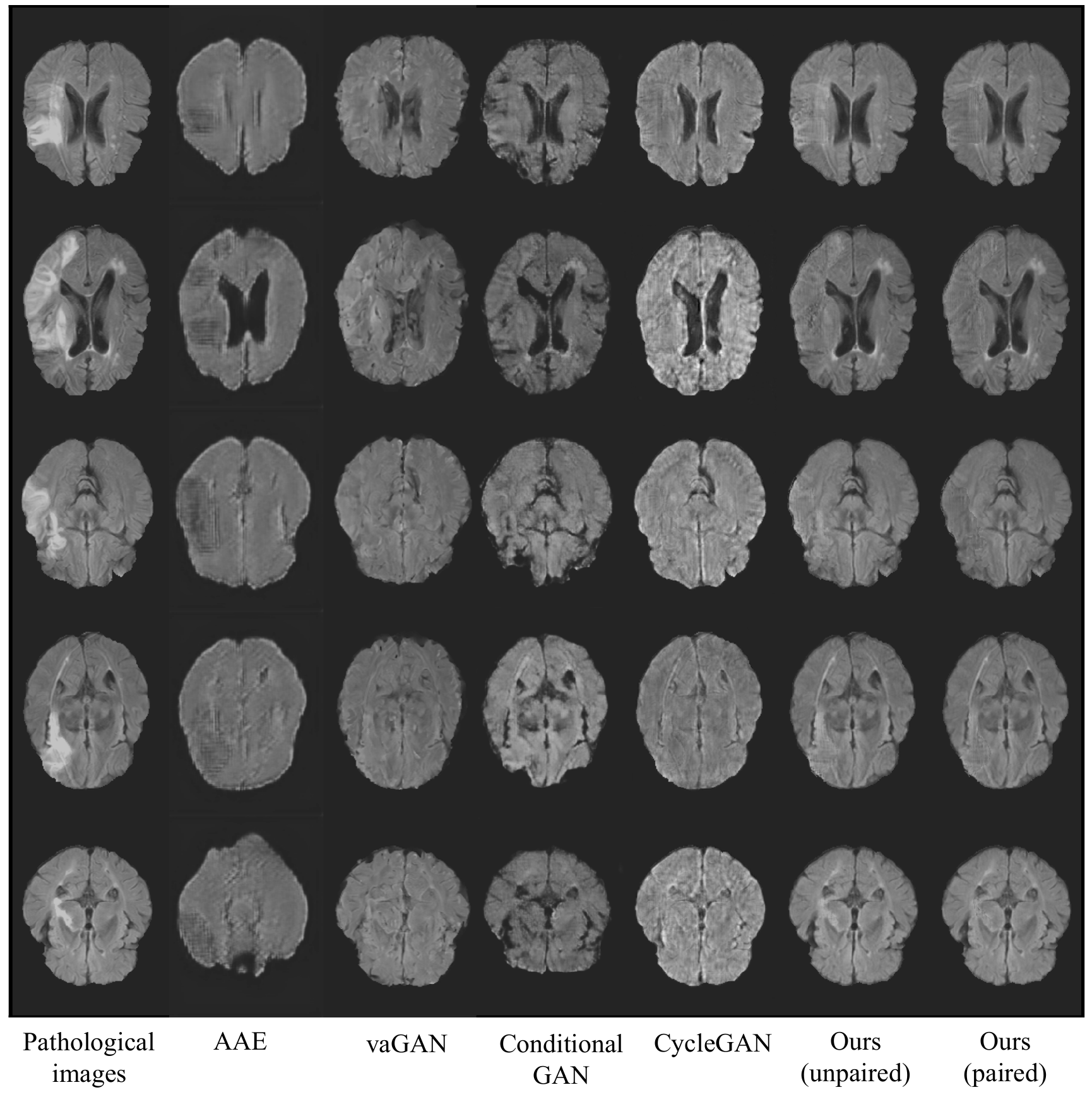}
    \caption{Experimental results of five samples (each in every row) for ISLES data.  The columns from left to right are the original pathological images, and the synthetic healthy images by \textit{AAE}, \textit{vaGAN}, \textit{Conditional GAN}, \textit{CycleGAN}, and the proposed method in the \textit{unpaired} and \textit{paired} setting, respectively.}
    \label{fig:isles_results}
\end{figure*}

Each panel was a montage of: input diseased image; ground truth segmentation mask; pseudo-healthy images obtained as outputs of the tested algorithms. The raters were blinded to which algorithm generated each image and image arrangement was randomised (for every panel shown).  The raters knew though that the first image was the input to the algorithms.      

Overall each rater reviewed 50 panels, each containing 6 images, with a score for 3 metrics, providing a total of 900 scores. Across the four raters 3600 scores were available.   We asked raters to limit time spent on a panel to be less than 3 minutes. 

\noindent \textbf{Real v.s. fake test:} As our approach focuses on image synthesis, we performed a human experiment where we requested raters to tell apart real from synthetic images. Specifically, we randomly selected 50 pathological slices, and used the methods discussed herein to generate corresponding pseudo-healthy images. As a result, we generated 300 images in total. Then, we randomly selected 300 real healthy images, and presented all images in a random order to four researchers who classified them as real or fake. We used a standardised viewing setting (screen size, distance from screen, illumination, monitor brightness) and limited evaluation time to 1 minute per image, and measured `realness' as the ratio of images labelled `real'.

\section{Results and discussion}
\label{sec: reults}
 
All results reflect testing sets and we report both averages and standard deviation. We use bold font to denote the best performing method (for each metric) and an asterisk (*) to denote statistical significance compared to the best performing baseline or comparison method (to keep in check multiple comparisons).  We use a simple paired t-test to test the null hypothesis that there is no difference between our methods and the best performing baseline, at the significance level of $5\%$. We found that differences are normally distributed in the quantitative metrics based on the D'Agostino and Pearson's normality test \citep{d1971omnibus,d1973tests}).


\subsection{Pseudo-healthy synthesis for ischemic lesions}
\label{sec_5_1}
\begin{table}[!t]
\centering
\caption{Numerical evaluation of our method and baselines on ISLES dataset in terms of \textit{identity} $iD$ and \textit{healthiness} $h$. For each metric, 1 is the best and 0 is the worst.  
The best mean values are shown in \textbf{bold}. Statistical significant results (5\% level) of our methods compared to the best baseline are marked with an asterisk (*).  }
\begin{tabular}{r|cc|cc}
\hline
\multicolumn{1}{c|}{\multirow{2}{*}{Method}} & \multicolumn{2}{c|}{T2} & \multicolumn{2}{c}{FLAIR} \\ 
 \cline{2-5} 
\multicolumn{1}{c|}{}  & $iD$  & $h$ & $iD$   & $h$  \\ \hline
AAE & $0.63_{0.07}$     & $0.71_{ 0.14 }$        & $0.66_{ 0.06}$       & $0.81_{ 0.09}$         \\
vaGAN  &   $ 0.72_{ 0.05}$       &     $0.77_{0.11 }$        & $0.75_{0.04}$       & $0.85_{0.08}$         \\
Cond. GAN  &     $0.75_{0.06}$      &        $0.74_{ 0.12}$     & $0.73_{0.05}$       & $0.83_{0.12}$         \\
CycleGAN  &        $0.82_{ 0.04}$   &     $0.76_{0.11}$        & $0.83_{ 0.05} $       & $0.81_{ 0.08}$         \\ \hline
Ours (unpaired) &    $0.93^{*}_{ 0.04}$     &       $0.84^{*}_{ 0.09}$     & $0.87_{ 0.04}$     & $0.88^{*}_{ 0.06}$         \\
Ours (paired)  &       $\mathbf{0.97}^{*}_{ 0.04}$    &         $\mathbf{0.85}^{*}_{0.08}$    & $\mathbf{0.94}^{*}_{ 0.03}$       & $\mathbf{0.89}^{*}_{ 0.07}$         \\ \hline
\end{tabular}
\label{tab: isles_results}
\end{table}

Here we perform pseudo-healthy synthesis on ISLES dataset, which contains diseased subjects with ischemic lesions. These lesions should not alter the brain's shape distal to the lesion much \citep{maier2017isles}, but rather manifest as hyper-intense regions in T2 and FLAIR modalities. As described in Section \ref{sec: data and preprocessing}, all methods are trained with a `healthy' set containing images that do not have an annotated lesion mask, and with a `pathological' set containing the remaining images. The exception is the AAE \citep{chen2018unsupervised}, which requires only `healthy' images for training. For our method in unpaired setting, we used approximately 100 masks from 3 subjects for training the mask discriminator. Standard spatial augmentations have been applied to prevent overfitting of the discriminator on the real masks. Note that the baseline and comparison methods do not require pathological masks for training.

We compare our method with the methods of Section \ref{sec:baselines} qualitatively and quantitatively. Numerical results of identity ($iD$) and healthiness ($h$), defined in Section \ref{sec:evaluation}, are summarised in Table \ref{tab: isles_results}, and examples of synthetic images are shown in Figure~\ref{fig:isles_results}.

In Table \ref{tab: isles_results} we can see  that our method trained in the paired setting achieves the best results, followed by our method trained in unpaired setting. Both paired and unpaired versions outperform all others. A key reason behind our methods' improved performance is the pathology disentanglement, which enables the accurate reconstruction of the input pathological images without hiding pathology information in the pseudo-healthy images. We can also observe from Figure \ref{fig:isles_results} that our methods produce sharp and lesion-free images, evidenced also by the superior healthiness values in Table \ref{tab: isles_results}. The synthetic images also preserve details of the input images, which points that subject identity is preserved along with image quality.

Furthermore, we observe (Table \ref{tab: isles_results}) that CycleGAN achieves the third best results in terms of \textit{identity}, which showcases the benefit of cycle-consistency loss in preserving subject identity. However, as described in Section \ref{sec:one_to_many_problem}, CycleGAN suffers from the \textit{one-to-many problem}, which misleads it to generate artifacts in synthetic images. As a result, the healthiness of CycleGAN is not as good as the ones of vaGAN and Conditional GAN, which do not need to `hide' pathology information in the pseudo-healthy images. 

Although vaGAN involves a $\ell_1$  loss between the input images and synthetic images, we do not see significant improvements over Conditional GAN, where such a regularization loss is not used. In Figure \ref{fig:isles_results}, we also observe a loss of subject {identity} in both {vaGAN} and {Conditional GAN}. Even though {vaGAN} produces results that maintain the outline of the brain, these results lack refined details. On the contrary, {Conditional GAN} changes the outline of the brain but maintains inner details.

In addition, {AAE}  often loses subject {identity}, and the produced synthetic images may present artifacts within the pathological areas of the input images. This is because there is no explicit loss to force the synthetic images to maintain the subject identity, neither a loss to explicitly ensure that the network learned to transform the pathological area to be `healthy'. 


\subsection{Pseudo-healthy synthesis for brain tumours}
\label{sec5_2}
\begin{figure}[t!]
    \centering
    \includegraphics[scale=0.30]{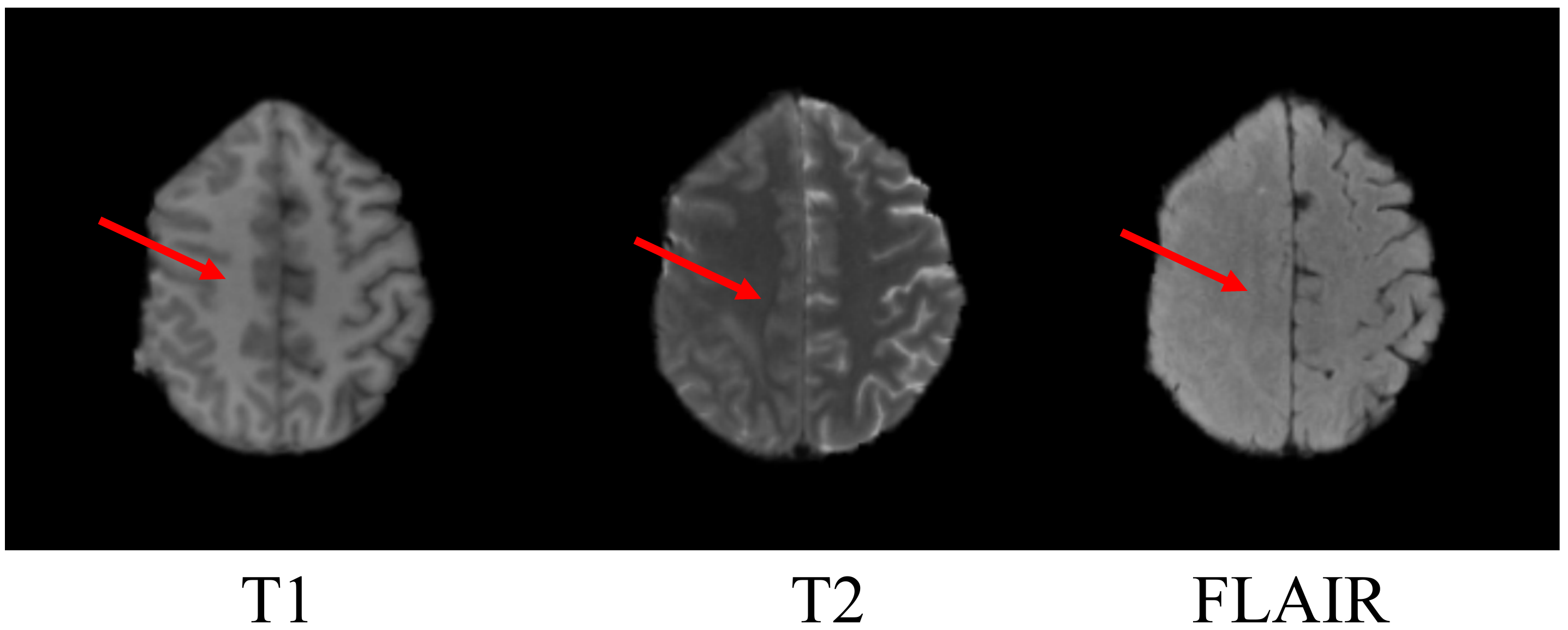}
    \caption{
    An example of BraTS images where glioblastoma is not present, but the brain tissues are still affected by deformations. From left to right are the same slice in T1, T2 and FLAIR modalities, respectively. The red arrows point to the affected areas, i.e. the left half of the brain.} 
    \label{fig: brats healthy deformation}
\end{figure}
\begin{figure*}[t!]
    \centering
    \includegraphics[scale=0.53]{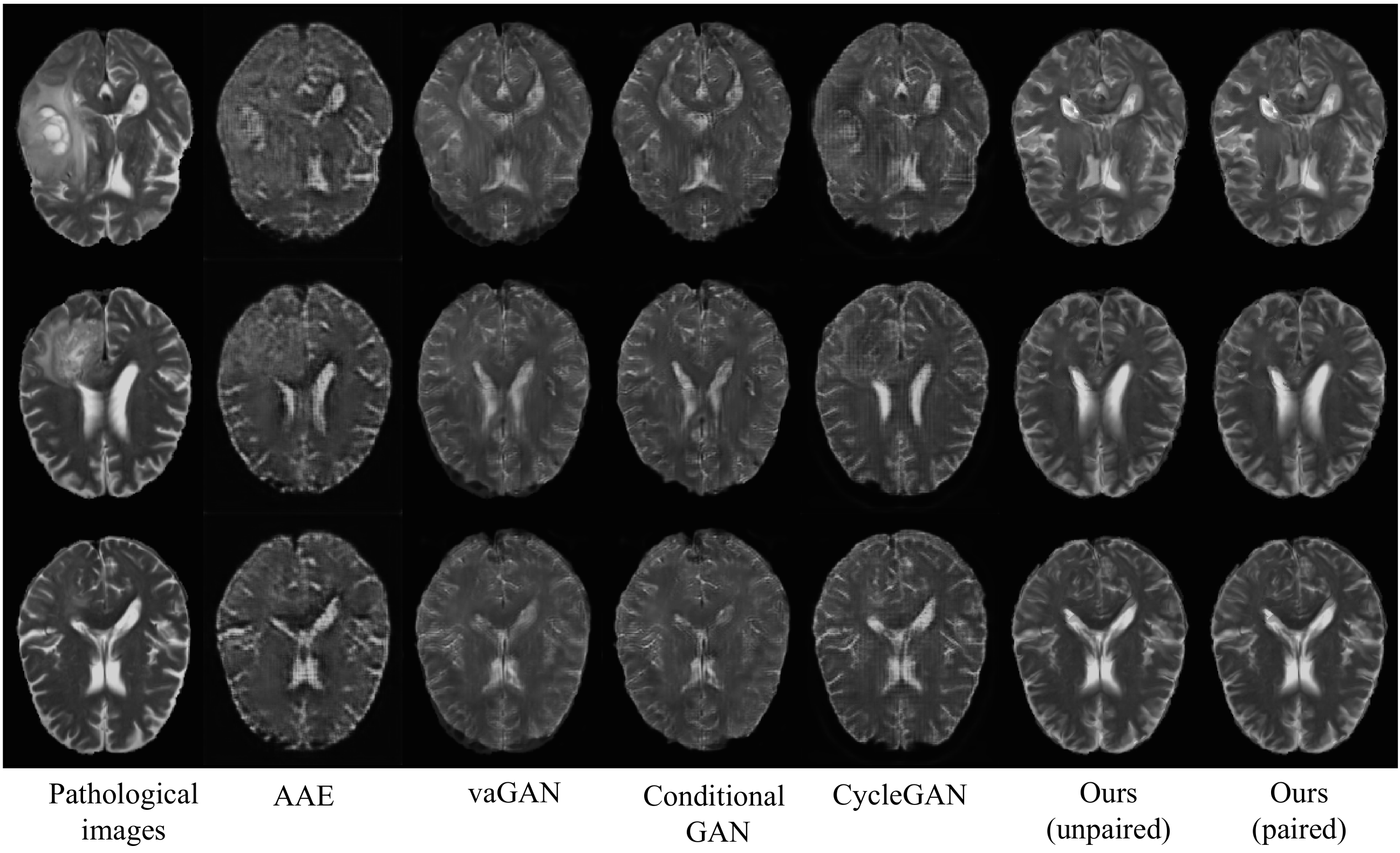}
    \caption{Experimental results of three samples, each in every row, for BraTS data.  The columns from left to right are the original pathological images, and the synthetic healthy images by \textit{AAE}, \textit{vaGAN}, \textit{Conditional GAN}, \textit{CycleGAN}, and the proposed method in the \textit{unpaired} and \textit{paired} setting, respectively.}
    \label{fig: deformation corrections}
\end{figure*}

\begin{table*}[t!]
\centering
\caption{
Results of our methods on BraTS dataset. Here we evaluate three metrics, defined in Section \ref{sec:evaluation} on T1 and T2 modalities.  For each metric, 1 is the best and 0 is the worst. We show also results (last three columns) of a human evaluation on the T2 modality based on criteria as described in Section \ref{sec:evaluation}. The best mean values are shown in \textbf{bold}. Statistical significant results (5 \% level) of our methods compared to the best baseline are marked with an asterisk (*). `def. corr.' is a shorthand for `deformation correction' assessment score from the raters.}

\begin{tabular}{r|ccc|ccc|ccc}
\hline
\multicolumn{1}{c|}{\multirow{2}{*}{Method}} & \multicolumn{3}{c|}{T1}                       & \multicolumn{3}{c|}{T2}                                             & \multicolumn{3}{c}{T2 (human evaluation)}     \\ \cline{2-10} 
\multicolumn{1}{c|}{}                        & $iD$          & $h$       & $DeC$        & $iD$          & $h$       & \multicolumn{1}{l|}{$DeC$}        & `identity'          & `healthiness'      & `def. corr.' \\ \hline
AAE                                          & $0.65_{0.12}$ & $0.72_{0.16}$ & $0.71_{0.05}$ & $0.63_{0.12}$ & $0.71_{0.13}$ & \multicolumn{1}{l|}{$0.75_{0.04}$} & $0.39_{0.34}$ & $0.30_{0.32}$ & $0.28_{0.31}$ \\
vaGAN                                        & $0.72_{0.11}$ & $0.79_{0.12}$ & $0.84_{0.06}$ & $0.74_{0.10}$ & $0.78_{0.09}$ & \multicolumn{1}{l|}{$0.81_{0.05}$} & $0.52_{0.34}$ & $0.49_{0.33}$ & $0.46_{0.39}$ \\
conditional GAN                              & $0.70_{0.14}$ & $0.73_{0.17}$ & $0.82_{0.04}$ & $0.69_{0.09}$ & $0.73_{0.15}$ & \multicolumn{1}{l|}{$0.84_{0.04}$} & $0.47_{0.32}$ & $0.46_{0.34}$ & $0.50_{0.31}$ \\
CycleGAN                                     & $0.82_{0.08}$ & $0.80_{0.13}$ & $0.71_{0.09}$ & $0.81_{0.07}$ & $0.77_{0.14}$ & \multicolumn{1}{l|}{$0.73_{0.06}$} & $0.56_{0.34}$ & $0.53_{0.35}$ & $0.30_{0.21}$ \\ \hline
Ours (unpaired)                          & $\textbf{0.84}_{0.08}$ & $0.82^{}_{0.11}$ & $\textbf{0.88}^{*}_{0.11}$ & $0.83_{0.06}$ & $0.83^{*}_{0.09}$ & \multicolumn{1}{l|}{$0.86^{*}_{0.05}$} & $0.65^{}_{0.29}$ & $0.67^{*}_{0.27}$ & $0.62^{*}_{0.25}$ \\
Ours (paired)                            & $0.83_{0.06}$ & $\textbf{0.86}^{*}_{0.10}$ & $0.85^{*}_{0.10}$ & $\textbf{0.85}^{*}_{0.04}$ & $\textbf{0.84}^{*}_{0.07}$ & \multicolumn{1}{l|}{$\textbf{0.88}^{*}_{0.04}$} & $\textbf{0.67}^{*}_{0.24}$ & $\textbf{0.69}^{*}_{0.23}$ & $\textbf{0.65}^{*}_{0.25}$ \\ \hline
\end{tabular}
\label{tab: brats deformation results}
\end{table*}

Here we apply our method on the BraTS dataset where volumes have high grade glioma. As described in Section \ref{sec_5_1}, for the case of ischemic lesions we used `healthy' images from the same dataset.  However, as shown in Figure \ref{fig: brats healthy deformation}, BraTS slices with no tumour annotations may still exhibit deformations. Furthermore, our previous work \citep{xia2019adversarial} showed that training with `healthy' slices from BraTS, only adjusted the intensities within the tumour areas, but was not able to fix the deformations caused by tumours.
We therefore use a second healthy dataset, Cam-CAN, to extract 2D healthy slices, which we used for model training, after confirming its suitability by comparing its intensity distribution with the one of BraTS (see Section~\ref{sec: data and preprocessing}). 
For our method in unpaired setting, and to train the mask discriminator, we used approximately 950 masks from 70 subjects that were not part of the training, validation and test sets. Standard spatial augmentations were applied to prevent overfitting of the discriminator on the real masks.



Figure \ref{fig: deformation corrections} shows visual comparisons between the methods considered. We observe that our method produces realistic results and preserves details, while other methods are more susceptible to losing subject identity. {CycleGAN} can better preserve identity, although image quality is deteriorated (see the bottom of the brain). In addition, {CycleGAN} creates some artifact inside the pathological region. It is possible that this artifact may indeed be the information that CycleGAN hides to enable input reconstruction. Furthermore, {Conditional GAN} and {vaGAN} produce images that are darker and do not match details of the input alluding to possible identity loss. This could be attributed to the lack of losses to help preserve identity, thus making it `easier' for {Conditional GAN} and {vaGAN} to learn a mapping from a pathological to a healthy image of a different subject.  Finally, {AAE} outputs appear  blurry and with visible artifacts inside the diseased region. 

Quantitative results are shown in Table \ref{tab: brats deformation results}, employing now three metrics including one that also assesses deformation correction, as previously described in Section~\ref{sec:evaluation}.
 
As expected, {identity} of our methods, as measured by $iD$, has dropped compared to Table \ref{tab: isles_results}. 
This is because our methods try to alter the structure of brains to fix the deformations. Indeed, when employing the new metric $DeC$, our methods achieve higher probability of generated images classified as `healthy'. 
For {healthiness}, $h$, our methods still outperform the other methods, indicating that the generated images do not contain detectable disease.

\subsection{Results of expert evaluation on pseudo-healthy synthesis for brain tumours}



In recognition that our metrics may partially reflect image quality as perceived by expert observers, herein we report the results of our observer study.  We aggregated the scores for each approach and averaged across raters to obtain a single consensus score per method per image, for which we used to calculate standard deviation and perform statistical analysis.  Given that categorical scores of the human raters and their differences are not normally distributed we instead use a bootstrapped paired t-test \citep{davison1997bootstrap} to test the null hypothesis described in Section \ref{sec_5_1}. 
 
The results of this analysis are shown in Table \ref{tab: brats deformation results}. We observe that our methods still outperform baselines and other methods, with a significant improvement for all metrics.
In addition, we observe that the methods ranking order is mostly preserved compared to the ranking obtained by the quantitative metrics. Intriguingly, CycleGAN can `fool' the pre-trained Segmentor which measures healthiness in the `h' metric but not expert observers in how they assess healthiness.
These observations suggest that while numerical   evaluation is generally consistent with expert evaluation, there can be room for improvement.  We note here the standard deviations for all methods are relatively high, which is due to the binary scoring system used for experiment. Furthermore, we obtained the point biserial correlation~\cite{tate1954correlation} between the values produced by our metrics and the human evaluation study to be 0.35, 0.32, and 0.36 for $iD$, $h$, and $DeC$, respectively. This implies a positive correlation between quantitative and human metrics.

To further evaluate the quality of synthesised images, we  requested human observers to discriminate between real and generated `healthy' images, as described in Section~\ref{sec:evaluation}. We calculated the `realness' score to be  $0.43\pm0.33$ for \textit{AAE}, $0.48\pm0.36$ for \textit{vaGAN}, $0.44\pm0.30$ for \textit{Conditional GAN}, $0.47\pm0.31$ for \textit{CycleGAN},  $0.51\pm0.31$ for our method (unpaired),  $0.54\pm0.25$ for our method (paired), and $0.63\pm0.32 $ for ground-truth healthy images as upper benchmark, .  Observe that our approaches were the closets to benchmarks.


\subsection{Segmentation results}

Here we evaluate the use of pseudo-healthy synthesis on segmentation of T2 BraTS images. Specifically, we compared the pseudo-healthy images with the ground-truth pathological images, and obtained the segmentation masks from the difference maps using a threshold of 0.1. For our method, and since segmentation is explicitly performed, we test with masks obtained both from the pseudo-healthy images, and from the \textit{Segmentor}.
We calculated Dice scores on the test sets to be $0.34\pm0.11$ for AAE, $0.53\pm0.13$ for vaGAN, $0.51\pm0.14$ for conditional GAN, and $0.63\pm0.16$ for CycleGAN. Our approach in unpaired setting obtained $0.74\pm0.14$ when using the \textit{Segmentor} output, and $0.70\pm0.13$ when using the pseudo-healthy images. In both cases our approach achieved statistically significant better results compared to the other benchmarks.

\subsection{Ablation studies}
\subsubsection{Semi-supervised learning}


In this section, we evaluate the effect of the amount of supervision by performing a semi-supervised experiment. Specifically, we vary the number of masks used in the supervised loss of Equation~\ref{eq: seg_paired}, while keeping the number of images fixed. The edge cases when all images have paired masks, and vice versa, correspond to the paired and unpaired setting respectively. Also, the number of segmentation masks used by the unsupervised loss of Equation~\ref{eq: seg_unpaired} is fixed in all cases.
The training strategy is that if the input image has a ground-truth pathology mask, then we use this mask to train the segmentor, with Equation~\ref{eq: losses in paired}. When the input image does not have a ground-truth pathology mask, we use the mask adversarial loss to train the network, with Equation~\ref{eq: losses in unpaired}. The results are presented in Table~\ref{tab: semi-supervised learning}.  
\begin{table*}[t]
\centering
\begin{tabular}{c|cccccc}
\hline
Ratio of \textit{paired} samples & 0\% (unpaired) & 20\%          & 40\%          & 60\%          & 80\%          & 100\% (paired) \\ \hline
$iD$                      & $0.87_{0.04}$  & $0.88_{0.05}$ & $0.90_{0.06}$ & $0.91_{0.05}$ & $0.93_{0.04}$ & $0.94_{0.03}$  \\
$h$                       & $0.88_{0.06}$  & $0.87_{0.06}$ & $0.89_{0.05}$ & $0.88_{0.06}$ & $0.89_{0.08}$ & $0.89_{0.07}$  \\ \hline
\end{tabular}
\caption{Numerical evaluation of our method on ISLES FLAIR dataset when the ratio of \textit{paired} samples changes. Here x\% means that x\% of the training pathological images have corresponding ground-truth pathology masks. }

\label{tab: semi-supervised learning}
\end{table*}

We can observe that for all paired sample ratios, our method can achieve synthetic images of great quality in terms of \textit{identity} and \textit{healthiness}. Nevertheless, we can observe that the \textit{iD}, i.e. identity score, increases as the ratio of the paired samples also increases. This could be attributed to the effect of more stable training of Segmentor. For ISLES dataset, the Generator needs to learn an identity mapping for  healthy regions and a pseudo-healthy function for pathological regions. The Segmentor performance has a direct effect on the Reconstructor and an indirect effect on the  Generator through back-propagation. With less supervision, the training of Segmentor is noisier, and the segmented pathological region, that Generator and Reconstructor focus on, is also noisier. Therefore, learning an identity and pseudo-healthy function is harder. This affects the identity score, as the Generator must learn to synthesise a whole brain image, and cannot reliably learn an identity function for some parts. On the contrary, the healthiness score, which is directly punished by the adversarial training loss, is not significantly affected. Finally, in order to perform a fair comparison, we trained models at a fixed number of epochs. Even though all models have converged, the noisier training due to the smaller amount of supervision have resulted in a different optimum and therefore to the drop of the identity metric. 

\subsubsection{Unsupervised segmentation and importance of cycle-consistency loss}

A pre-requisite for an accurate pseudo-healthy synthesis that does not contain traces of pathological information, is for the Segmentor \textit{S} to be able to accurately extract masks, such that they can be used for the reconstruction of the input pathological images. This should be possible in the {unpaired} setting as well, where the 
{Segmentor} is not trained with any supervision cost. In this setting, the {Segmentor} is trained using the adversarial loss of the mask discriminator (Equation \ref{eq: seg_unpaired}), as well as the cycle-consistency loss (Equation \ref{eq: loss_cc1}) of the input images.

We evaluate the accuracy of \textit{S} in the {paired} and {unpaired}  setting on FLAIR images from ISLES: we obtain respectively an average Dice score of $0.87\; (0.15)$ and $0.79\; (0.17)$ in the testing sets. The results show that even in the {unpaired} setting, our method can still achieve good segmentation. Results appear to be on par with the numbers provided in \citet{andermatt2018pathology}.
To demonstrate the importance of the cycle-consistency loss (Equation \ref{eq: loss_cc1}), we perform an ablation study where we train \textit{S} only with the adversarial loss of the mask discriminator (i.e.\ only  with Equation \ref{eq: seg_unpaired}).  We found that this achieves  a Dice of $0.66\; (0.19)$ which is much lower than before. This highlights that just matching the adversary is not enough and that the cycle-consistency loss, by backpropagating additional gradients to the segmentor originating from this cost, encourages further the segmented mask to be correct (in place and size) to enable better reconstruction of the input pathological image.

\subsubsection{Usefulness and design of Cycle H-H}

Our method includes a second training cycle, Cycle H-H, that reconstructs healthy images and masks. This cycle improves the identity preservation of the input images and ensures that our method does not invent disease when a healthy image is given. 

Here we perform two ablation studies. For the first ablation study, we train our methods without {Cycle H-H}, i.e. train the network only with {Cycle P-H}. For the second ablation study, we change Cycle H-H to a new cycle, termed {Cycle H-P}, which translates healthy images to synthetic diseased ones. The difference between {Cycle H-H} and {Cycle H-P} is that {Cycle H-H} translates a healthy image and a healthy mask to a fake healthy one, and then reconstructs the input healthy image and mask; while {Cycle H-P} translates a healthy image and a pathology mask to a fake diseased one, and then reconstructs the input healthy image and pathology mask.  The training of {Cycle H-P} requires an additional discriminator to encourage realistic synthesis of pathological images, and requires careful selection of pathology masks that are suitable to guide the pseudo diseased image generation and fit the real healthy images. We perform the experiments in {paired} setting on ISLES FLAIR images. 

The results are shown in Table \ref{tab: ablation on cycle h-h}. We observe that our method with {Cycle H-H} outperforms variants without it and with {Cycle H-P}. This highlights the importance and effectiveness of the simple, yet effective, design of {Cycle H-H} in preserving subject {identity} and improved {healthiness} of pseudo-healthy images.

\subsubsection{Effectiveness of Wasserstein loss}
In this paper, to train the discriminators, we replaced the LS-GAN loss~\citep{mao2017least} that we used previously~\citep{xia2019adversarial}, with the Wasserstein loss with gradient penalty~\citep{gulrajani2017improved}, which we found to further stabilise training and improve the generated image quality.  To illustrate the latter, in Table~\ref{tab: ablation on cycle h-h} we also show results from models trained in the paired setting on ISLES FLAIR images when using the LS-GAN loss. We observe that Wasserstein loss improves quantitatively the synthetic images in terms of identity and healthiness.

\begin{table}[t!]
\centering
\caption{Ablation studies. Here we compare our model with ablated models where we train in the paired setting on ISLES: without Cycle H-H; train with a modified \textit{Cycle H-P} cycle; and also train with Least Square discriminator loss. See text for more details.}
\begin{tabular}{c|cc}

\hline
Method               & iD & h \\ \hline
without Cycle H-H     &     $0.85_{0.05}$        &   $0.93_{0.04}$       \\
With Cycle P-H        &      $0.89_{0.06}$       &    $0.89_{0.04}$      \\ 
With LS-GAN loss & $0.92_{0.03}$ & $0.97_{0.04}$  \\\hline
Ours  (Cycle H-H \& Wasserstein) &     $0.94_{0.03}$        &      $0.99_{0.03}$    \\ \hline
\end{tabular}
\label{tab: ablation on cycle h-h}
\end{table}






\subsubsection{Pseudo disease synthesis}

\begin{figure*}[t]
    \centering
    \includegraphics[scale=0.53]{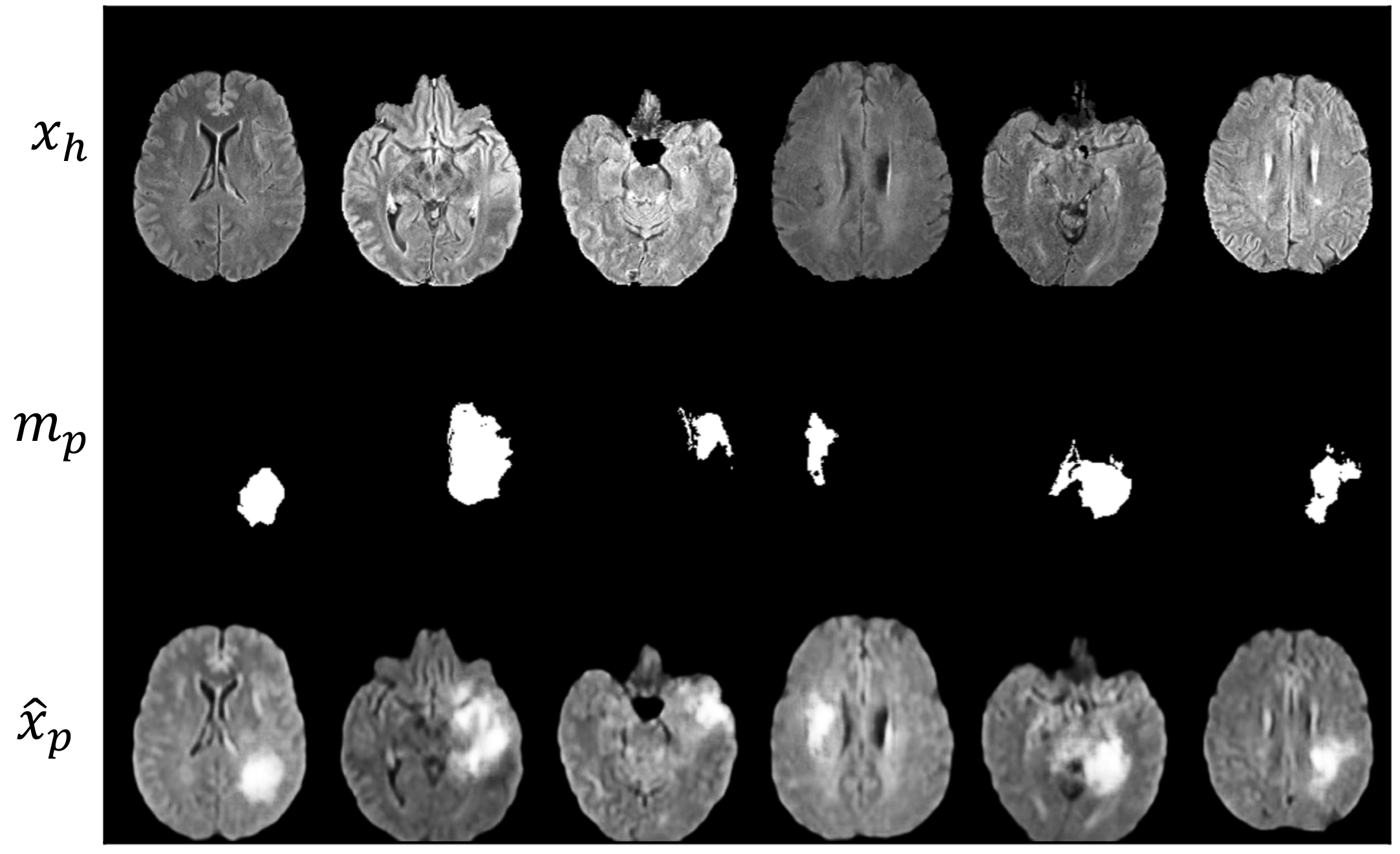}
    \caption{Pseudo disease synthesis. Top row shows healthy images, middle row shows random pathology masks, and bottom row presents the synthetic `pathological' image by the Reconstructor. We can see that Reconstructor can generate realistic 'pathological' images based on input images and masks.}
    \label{fig: pseudo disease synthesis}
\end{figure*}

If our method works well, the Reconstructor should be able to synthesise a `pathological' image given a healthy one and a suitable pathology mask. Here we used randomly sampled but suitable masks and healthy images to generate a pathological image, with a trained Reconstructor on ISLES FLAIR dataset. We show some example images of this pseudo disease synthesis, as shown in Figure \ref{fig: pseudo disease synthesis}. We can observe that although our model has never been trained to perform this pseudo disease synthesis, the Reconstructor is still able to synthesise a `pathological' image when given a healthy image and a suitable pathology mask. 

\section{Conclusion}
\label{sec6}

We presented a method that aims to synthesise pseudo-healthy images using an adversarial design that disentangles pathology. Our method is composed of a {Generator} that creates pseudo-healthy images and a  {Segmentor} that predicts a pathology map.  These key components are trained aided by  the {Reconstructor}, which reconstructs the input pathological image conditioned on the map and the pseudo-healthy image. Our method can be trained using supervised and adversarial loses taking advantage  of unpaired data. We propose numerical evaluation metrics to explicitly measure the quality of the synthesised images. We demonstrate on ISLES,  BraTS and Cam-CAN datasets that our method outperforms baselines both qualitatively, quantitatively, and subjectively with a human study.

We see several avenues for future consideration by us or the community at large. Metrics that enforce or even measure identity is a topic of considerable interest in computer vision \citep{antipov2017face}. One of our proposed metrics aimed to assess whether the subject {identity} has been preserved in synthetic `healthy' images, while another metric assessed if deformation caused by disease was  recovered. Analysis combining these two metrics could assess the preservation of {identity} even when deformation was corrected which is suited for cases where disease globally affects an image. Further lines of improvement involve better methods to measure the null hypothesis (e.g.\ perhaps by artificially creating images from the healthy class that seem to be distorted).  In addition, we do see that human evaluation is useful, although challenging since it requires expertise. Moreover, most clinical neurologists do not evaluate medical images in isolation, but rather consider them in combination with other medical information, in order to make a diagnostic decision. Nevertheless, we have performed a human experiment involving a neurologist, which best adhered to a blinded workflow. However, better evaluation schemes could be proposed which is seen as a future direction. We also see a future opportunity in creating a large benchmark study that amasses expert evaluations which are used to learn combinations of several quantitative, yet easy to obtain, numerical metrics that can act as surrogates to human evaluations. Furthermore, extending this work to disentangle different factors, such as multiple diseases, could explain for example their effect on the brain, and thus characterise the severity of each one. Finally, this method despite our efforts to introduce 3D networks  remains 2D: we found the parameter space (and GPU memory) exploding due to the several networks. Finally, many datasets are multimodal so there could be a benefit in creating multi-input multi-output models; however, this may necessitate different generators (one per modality) further increasing parameter space.

\section*{Acknowledgments}
This work was supported by the University of Edinburgh by PhD studentships to T. Xia and A. Chartsias and used resources provided by the Edinburgh Compute and Data Facility
(http://www.ecdf.ed.ac.uk/). This work was partially supported by EPSRC (EP/P022928/1) and by The Alan Turing Institute under the EPSRC grant EP/N510129/1. This work was  supported in part by the US National Institutes of Health (R01HL136578). We thank Nvidia for donating a Titan-X GPU.  S.A. Tsaftaris acknowledges the support of the Royal Academy of Engineering and the Research Chairs and Senior Research Fellowships scheme. We wish to thank members of the lab (Dr. Valerio Giuffrida, Dr. Haochuan Jiang, Dr. Spyridon Thermos, Mr. Grzegorz Jacenkow, Mr. Gabriele Valvano and Mr. Xiao Liu  )  and an anonymous expert for helping to visually assess the images. Finally, we thank Dr. Dafan Yu, a clinical neurologist with the Third Affiliated Hospital of Sun Yat-sen University, for helping to evaluate the results.

\bibliographystyle{model2-names.bst}\biboptions{authoryear}
\bibliography{refs}

\end{document}